\documentclass[twocolumn,apj,iop]{openjournal}
\usepackage{hyperref}

\hypersetup{
    colorlinks=true,
    linkcolor=blue,
    filecolor=magenta,      
    urlcolor=cyan,
}

\usepackage{xspace}
\usepackage{float}
\usepackage[caption=false]{subfig}
\usepackage[T1]{fontenc}
\usepackage{graphicx}% Include figure files
\usepackage{dcolumn}% Align table columns on decimal point
\usepackage{bm}% bold math
\let\tablenum\relax
\usepackage[input-decimal-markers=.]{siunitx}
\usepackage{lipsum}

\mathchardef\mhyphen="2D

\begin{document}
%\nolinenumbers
\preprint{APS/123-QED}

\title{Dissecting the Thermal SZ Power Spectrum by Halo Mass and Redshift in SPT-SZ Data and Simulations}% Force line breaks with \\

\author{Josemanuel Hern\'andez} 
\affiliation{Department of Astronomy \& Astrophysics, The University of Chicago, Chicago, IL 60637, USA}
\affiliation{Kavli Institute for Cosmological Physics, The University of Chicago, Chicago, IL 60637, USA}
\author{Lindsey Bleem}
\affiliation{High-Energy Physics Division, Argonne National Laboratory, 9700 South Cass Avenue., Lemont, IL, 60439, USA}
\affiliation{Kavli Institute for Cosmological Physics, The University of Chicago, Chicago, IL 60637, USA}
\author{Thomas Crawford}
\affiliation{Department of Astronomy \& Astrophysics, The University of Chicago, Chicago, IL 60637, USA}
\affiliation{Kavli Institute for Cosmological Physics, The University of Chicago, Chicago, IL 60637, USA}
\author{Nicholas Huang}
\affiliation{Department of Physics, University of California, Berkeley, CA, 94720, USA}
\author{Yuuki Omori}
\affiliation{Department of Astronomy \& Astrophysics, The University of Chicago, Chicago, IL 60637, USA}
\affiliation{Kavli Institute for Cosmological Physics, The University of Chicago, Chicago, IL 60637, USA}
\author{Srinivasan Raghunathan}
\affiliation{Center for AstroPhysical Surveys, National Center for Supercomputing Applications, Urbana, IL, 61801, USA}
\author{Christian Reichardt}
\affiliation{School of Physics, University of Melbourne, Parkville, VIC 3010, Australia}
%\author{SPT Collaboration}
%\affiliation{
% University of Chicago,\\
% The South Pole Telescope Project% %with \\
%}%

\begin{abstract}

We explore the relationship between the thermal Sunyaev-Zel'dovich (tSZ) power spectrum amplitude and the halo mass and redshift of galaxy clusters in South Pole Telescope (SPT) data, in comparison with three $N$-body simulations combined with semi-analytical gas models of the intra-cluster medium. Specifically, we calculate both the raw and fractional power contribution to the full tSZ power spectrum amplitude at $\ell = 3000$ from clusters as a function of halo mass and redshift. 
We use nine mass bins in the range $1 \times 10^{14}\ M_\odot\ h^{-1} < M_{500} < 2 \times 10^{15}\ M_\odot\ h^{-1}$, and two redshift bins defined by $0.25 < z < 0.59$ and $0.59 < z < 1.5$. We additionally divide the raw power contribution in each mass bin by the number of clusters in that bin, as a metric for comparison of different gas models. At lower masses, the SPT data prefers a model that includes a mass-dependent bound gas fraction component and relatively high levels of AGN feedback, whereas at higher masses there is a preference for a model with a lower amount of feedback and a complete lack of non-thermal pressure support. The former provides the best fit to the data overall, in regards to all metrics for comparison. Still, discrepancies exist and the data notably exhibits a steep mass-dependence which all of the simulations fail to reproduce. This suggests the need for additional mass- and redshift-dependent adjustments to the gas models of each simulation, or the potential presence of contamination in the data at halo masses below the detection threshold of SPT-SZ. Furthermore, the data does not demonstrate significant redshift evolution in the per-cluster tSZ power spectrum contribution, in contrast to self-similar model predictions.
\end{abstract}

\keywords{Large-Scale Structure of the Universe, Galaxy Clusters, Intracluster Medium, Astronomical Simulations, Sunyaev-Zeldovich Effect}
%\correspondingauthor{Josemanuel Hernandez}
%\email{chema.hernandez.377@gmail.com}

\section{Introduction}
\label{sec:intro}

Galaxy clusters are the largest gravitationally bound structures in the Universe, and are principally composed of dark matter, galaxies (stars), and ionized gas. As a result of ionization, these structures are hosts to a reservoir of hot free electrons which reside in what is known as the intra-cluster medium (ICM). The inverse Compton scattering of cosmic microwave background (CMB) photons off of these hot electrons is known as the thermal Sunyaev Zel’dovich (tSZ) effect  \citep{1970Ap&SS...7....3S}. Since the strength of the tSZ signal is redshift-independent, it enables the study of clusters across a wide range of cosmic time. The tSZ signal is a direct probe of the pressure profiles of galaxy clusters, which makes it a compelling probe of both cluster physics and cosmology \citep{Carlstrom_2002}.

The angular power spectrum of the tSZ effect is an especially sensitive probe of structure evolution and can be used to constrain the value of $\sigma_{8}$, the cosmological parameter that describes the amplitude of the matter power spectrum at a scale length of 8 Mpc $h^{-1}$. The tSZ power spectrum amplitude is predicted to scale as $\sim \sigma^{7-8}_{8}$ \citep{Komatsu_2002}.  

A computationally efficient way to simulate and model the tSZ is through the use of high-volume $N$-body (dark matter only) simulations in combination with semi-analytical ICM gas models. These semi-analytical gas models can be calibrated against low-volume hydrodynamical simulations, and data, to yield high-accuracy and high-volume tSZ simulations. Semi-analytical modelling of the ICM gas often involves calibration against X-ray data specifically. Yet, there is a lack of X-ray data for the high-redshift low-mass clusters which principally source the tSZ power spectrum at small angular scales \citep{Komatsu_2002}, which leads to major modelling limitations. In this paper we consider three such simulations, in comparison with data from the 2500 deg$^2$ SPT-SZ survey conducted with the South Pole Telescope (SPT). Specifically, we analyze the amplitude of the tSZ power spectra at small scales (high multipole number) in these data and simulations as a function of cluster halo mass. Redshift bins are also employed as a probe of redshift-evolution. We note that this analysis is similar in spirit to \cite{2021MNRAS.503.5310R}, but with an explicit focus on mass and redshift values. We are motivated by the potential to provide further constraints on models and simulations of the tSZ.

This paper is organized as follows. We discuss theoretical background on the tSZ effect in Section \ref{sec:TSZE}. In Section \ref{sec:sims} we describe the three tSZ simulations considered here and compare their different ICM gas models. We describe the data products used in Section \ref{sec:data}. In Section \ref{sec:methods} we outline the methods implemented. Our baseline results are presented in Section \ref{sec:baseline_results}, with a summary and discussion presented in Section \ref{sec:summary_and_discussion}. In Section \ref{sec:robustness_tests} we describe tests for evaluating the robustness of our baseline results. Finally, we conclude in Section \ref{sec:conclusion}.

\section{Background on the Thermal SZ Effect}
\label{sec:TSZE}
The tSZ effect is caused by the inverse Compton scattering interaction between hot ICM electrons and the CMB, which results in some CMB photons gaining energy and shifting to higher frequencies. The non-relativistic tSZ signal has a null frequency of 218 GHz. At frequencies below this, the tSZ signal can be observed as an intensity decrement in the CMB. At frequencies above this, it shows up as an intensity increment (e.g., \citealt{Carlstrom_2002}).

These distortions can be modeled as a change in the apparent CMB temperature, at a frequency of $\nu$ and at a particular direction on the sky $\hat{n}$, using the following equation for the non-relativistic limit:
\begin{equation}
    \frac{\Delta T_{\rm tSZ}}{T_{\rm CMB}} (\hat{n})  = f_{\nu}y(\hat{n}) 
\label{eq:1}
\end{equation}
where $f_{\nu}$ is a frequency dependent function, defined as:
\begin{equation}
    f_{\nu} = x{\rm coth}\left(\frac{x}{2}\right)-4,\
\label{eq:3}
\end{equation}
with
\begin{equation}
    x = \frac{h{\nu}}{k_{\rm B}T_{\rm CMB}},
\label{eq:4}
\end{equation} 
and $y$ is the Compton-$y$ parameter, defined as: 
\begin{equation}
    y(\hat{n})  = \frac{k_{\rm B}{\sigma}_{\rm T}}{m_{\rm e}c^{2}} \int n_{\rm e} (\vec{x}) T_{\rm e} (\vec{x}) dl(\hat{n}),
\label{eq:2}
\end{equation}
where $k_{\rm B}$ is the Boltzmann constant, $\sigma_{\rm T}$ is the Thomson cross-section, $m_{\rm e}$ is the electron mass, $n_{\rm e}$ is the electron number density, $T_{\rm e}$ is the electron temperature, and the integral is along the line of sight.

 The Compton-$y$ parameter is proportional to the integral of electron pressure along the line of sight. The description of the electron pressure is one of the main sources of of discrepancy between different models and simulations, and reflects an incomplete understanding of cluster gas physics.

The tSZ power spectrum is commonly reported in terms of $D_{\ell}$, where 
\begin{equation}
    D_{\ell} = \frac{\ell (\ell + 1)}{2 \pi} C_{\ell},
\label{eq:D_ell}
\end{equation}
and $C_\ell$ is the angular power spectrum, i.e., the variance of the coefficients $a_{\ell m}$ of the spherical harmonic transform of a sky map (or the covariance of the transforms of two sky maps), averaged over azimuthal index $m$.
In this work we take auto-spectra of simulated Compton-$y$ maps and denote this by $D^{yy}_{\ell}$. For the data, we take the cross-spectra of two half Compton-$y$ maps (${\rm half}1 \times {\rm half}2$), each created using half of the total data set from \citet{Bleem_2022}, and also denote this by $D^{yy}_{\ell}$ (see Section \ref{sec:data} for more details on the data).  

\section{Simulations}

In this section we describe three $N$-body simulations that are coupled with semi-analytical ICM gas models to simulate the tSZ signal. We present these products in the chronological order in which they were released and give brief overviews of the cosmology and gas physics employed in each. Table \ref{astro_quantities_description} gives a qualitative summary of some of the relevant astrophysical quantities in each simulation. We direct the reader to the associated papers of each product for more detailed descriptions.

%\dhayaa{Somewhere in this section, it would be good to have a table that sorta qualitatively distinguishes the different sims. So basically like Table 1, but it focuses on things like Nonthermal pressure support (does it exist or not, it has stronger scaling with redshift than the other sims? etc), AGN feedback (stronger, weaker etc.), self-similarity, bound gas fractions etc... I think that will help organize how the different sims compare with one another in terms of their predictions and astrophysical modelling choices, at least according to what we know from existing works.} \chema{See new Table \ref{astro_quantities_description}.}

\label{sec:sims}
\subsection{Sehgal et al. (2010) Microwave Sky Simulation}
The microwave sky simulation described in \citet{Sehgal_2010}, hereafter referred to as S10, is a publicly available $N$-body simulation with $1024^3$ dark matter particles in a 1000 ${\rm Mpc}\ h^{-1}$ box and cluster halos identified via a friends-of-friends (FOF) algorithm. The simulation is combined with a semi-analytic ICM gas model from \citet{Bode_2009}, which is calibrated against X-ray data \citep{Vikhlinin_2006,2009ApJ...693.1142S}, to produce simulated tSZ maps at six different frequencies. Other microwave sources are also simulated (including dusty star forming galaxies, active galactic nuclei, and the primary CMB), but are of lesser importance for the purposes of our analysis. 

Table \ref{table:cosmoparam} describes the cosmological parameters used by S10, along with those of two other simulations. For the cosmology adopted, S10 compares their simulated halo abundance to the semi-analytic fitting formula for the halo mass function from \citet{10.1046/j.1365-8711.2001.04029.x} and finds good agreement. 

The maps produced through this simulation are full-sky HEALPix (Hierarchical Equal Area isoLatitude Pixelization, \citealt{2005ApJ...622..759G}) maps of 0.4' resolution (HEALPix $N_{\rm side} = 8192$). To produce full-sky maps, S10 generates a halo catalog for a single octant and then replicates it seven more times to cover the full-sky area. This is then used to create full-sky maps of each separate microwave component as well as a combined sky map, at each of the six frequencies. The tSZ power spectrum obtained from these maps has been observed to have a higher amplitude than that obtained from data \citep{George_2015} and other simulations (see Fig.~\ref{fig:all_sim_tsz_ps}). This is likely due, in large part, to the absence of non-thermal pressure support in S10 (\citealt{Shaw_2010}; \citealt{osti_21567548}; \citealt{Battaglia_2012}). A rescaled Compton-$y$ map was produced in 2019, using a scaling factor of 0.75, to bring the amplitude of the tSZ power spectrum into good agreement with the 2013 {\it Planck} Compton-$y$ map power spectrum \citep{Ade_2019}. However, since we are interested in analyzing the effects of specific gas prescriptions, we use the original unscaled maps. We exclusively make use of the 148 GHz tSZ map and apply a frequency-dependent conversion factor (see Equation \ref{eq:3}) to convert it into units of Compton-$y$.

\begin{table}
\centering
\begin{tabular}[t]{cccc}
\toprule
Parameters & S10 & O22 & Agora\\
\hline
$\sigma_{8}$&0.8&0.82&0.818\\
$\Omega_{\rm b}$&0.044&0.046&0.0482\\
$\Omega_{\rm m}$&0.264&0.279&0.307\\
$\Omega_{\Lambda}$&0.736&0.721&0.693\\
$h$&0.71&0.7&0.678\\
$n_{\rm s}$&0.96&0.97&0.96\\
\hline
\end{tabular}
\caption{Cosmological parameters and values adopted by the three simulations discussed in this work.}\label{table:cosmoparam}
\end{table}

\subsubsection{S10 ICM Gas Model Prescription}

To simulate the tSZ signal, S10 adds a semi-analytical ICM gas prescription, which assumes a polytropic equation of state and hydrostatic equilibrium, to the $N$-body halos. Three separate approaches to modeling the gas are taken, depending on halo mass and redshift. For halos with FOF mass above $3 \times 10^{13}\ M_\odot\ $ and $z<3$, the model described in \citet{Bode_2009} is used. This category of halos is the main focus of the analysis presented here. Less massive and higher-redshift halos are more difficult to model because of a lack of X-ray data. For less-massive (down to $\sim1 \times 10^{13} \ M_\odot$) low-redshift ($z<3$) halos, S10 assigns a pressure based on the simulated velocity dispersion of a halo particle and its neighbors. For high-redshift halos ($3 <z<10$), mass density and momentum shells at various redshift slices in an isothermal universe are used to construct the tSZ signal. 

The \citet{Bode_2009} model includes four free parameters, calibrated against X-ray gas fractions as a function of temperature from Chandra X-ray observatory data of the time \citep{Vikhlinin_2006,2009ApJ...693.1142S}. The four parameters include: star formation rate, AGN feedback, dynamical energy transfer from dark matter to gas, and non-thermal pressure support. It is this last parameter, the amount of non-thermal pressure support, which is the principal difference between this model and the others described in this paper. \citet{Bode_2009} and S10 adopt a value of zero for the only parameter that describes the amount of non-thermal pressure support from merger-induced shocks, consequently yielding a higher value for the amplitude of the tSZ power spectrum \citep{Shaw_2010}. This is because non-thermal pressure support tends to reduce the tSZ signal at the outskirts of galaxy clusters \citep{osti_22086550}.

Table \ref{table:gasparam} describes the gas parameter values used to simulate the tSZ signal in S10, as well as that of one other simulation analyzed in this work. S10 yields a tSZ power spectrum amplitude of $D^{yy}_{3000} = 1.21 \times 10^{-12}$ in units of Compton-$y$ at $\ell = 3000$. Figure \ref{fig:all_sim_tsz_ps} shows the S10 tSZ power spectrum for a wider range of $\ell$ values, along with the power spectra of two other simulations and a data value from \cite{George_2015}.

\subsection{Osato \& Nagai (2022) Simulation}

\citet{https://doi.org/10.48550/arxiv.2201.02632} take realizations of an $N$-body simulation from \citet{Takahashi_2017}, along with their halo catalogs, and apply a halo-based pasting algorithm to simulate the tSZ signal. The \citet{Takahashi_2017} simulation is an $N$-body simulation created using six independent sets of 14 nested periodic boxes, ranging in side length from $450\ {\rm Mpc}\ h^{-1}$ to $6300\ {\rm Mpc}\ h^{-1}$, and increased in increments of $450\ {\rm Mpc}\ h^{-1}$. Each of these nested boxes contains $2048^{3}$ dark matter particles, which are evolved into dark matter halos at different redshifts. These are all run with $L-Gadget2$ code \citep{2005Natur.435..629S}, with halos identified using the $Rockstar$ algorithm \citep{2013ApJ...762..109B}. Table \ref{table:cosmoparam} lists the specific cosmological parameter values used by \citet{Takahashi_2017}. More details about the methods used can be found in the associated paper for this simulation. From now on we refer to the combination of the \cite{Takahashi_2017} $N$-body simulation and the \citet{https://doi.org/10.48550/arxiv.2201.02632} gas prescription as O22. We specifically only make use of one realization (realization000) of the 108 full-sky maps generated by O22.

\subsubsection{O22 ICM Gas Model Prescription}
In the O22 Compton-$y$ map used in this work, only the gas embedded within halos contributes to the tSZ signal. This means there is no contribution to the signal from any diffuse gas outside of the halo's virial radius. This is in contrast to a particle-based pasting algorithm, where both gas inside and outside the halo can contribute. O22 show that the tSZ power spectra of both the particle-based and halo-based pasting approaches agree reasonably well up to $\ell \sim 3000$ (see Figure 8 in the associated paper). The halo-based approach has the advantage of being more computationally efficient. 

O22 uses similar gas model parameters to those in S10, described in \citet{Bode_2009,Ostriker_2005}, and calibrated against X-ray gas fractions measured using data from Chandra (\citealt{Vikhlinin_2006,2009ApJ...693.1142S}) and XMM-Newton \citep{Lovisari_2015}. The parameters adopted by O22 are additionally calibrated against Chandra X-ray measurements of gas density profiles for $83$ massive SPT-SZ clusters \citep{2013ApJ...774...23M}. Furthermore, O22 also incorporates three distinct parameters to specify the amount of non-thermal pressure support in their model, as described and calibrated in \citet{Shaw_2010}. This is a major departure from the model of S10. \cite{Shaw_2010} finds that incorporating non-thermal pressure support in their gas model effectively scales the amplitude of the tSZ power spectrum down by roughly a factor of two at $\ell = 3000$ (See Figure 8 in \citealt{Shaw_2010}). Beyond non-thermal pressure support, all other gas model parameters adopted by O22 are those calibrated in \citet{2017ApJ...837..124F}. We list and describe all of these in Table \ref{table:gasparam}. 

O22 predicts a tSZ power spectrum amplitude of $D^{yy}_{3000} = 7.29 \times 10^{-13}$ in units of Compton-$y$ at $\ell = 3000$. The full tSZ power spectrum can be seen in Figure \ref{fig:all_sim_tsz_ps}. 

\subsection{Agora (2022) Simulation}
Following the approaches of S10 and O22, \cite{omori2022} applies analytical prescriptions to the MultiDark Planck 2 (MDPL2) $N$-body simulation to simulate different components of the microwave sky, including the tSZ effect. We will henceforth refer to the combination of MDPL2 and this tSZ gas model as Agora \citep{omori2022}.

MDPL2 is a dark matter-only $N$-body simulation of $3840^{3}$ particles in a box with a $1000\ h^{-1}{\rm Mpc}\ $ side length \citep{2016MNRAS.457.4340K}. Halos are identified by using the $Rockstar$ algorithm \citep{2013ApJ...762..109B}, in a similar fashion as O22. However, unlike O22, distinct lightcones are not generated by stacking a series of nested boxes with different resolutions. Instead, the original high-resolution box is tessellated to fill the simulation volume needed. This allows for a higher mass resolution at high redshifts.

The specific cosmological parameter values used by MDPL2 are listed in Table \ref{table:cosmoparam}. More details about the methods used can be found in \citet{2016MNRAS.457.4340K} and in \cite{omori2022}.

\subsubsection{Agora Gas Model Prescription}
\label{subsec:Agora_gas}
Agora uses the gas model described in \citet{Mead_2020}, which is based on the fitting of electron pressure profiles from the Baryons and Haloes of Massive Systems (BAHAMAS) hydrodynamical simulation \citep{2017MNRAS.465.2936M, McCarthy_2018}. The gas model of \cite{Mead_2020} employs a distinct set of parameters, different from those adopted by S10 and O22. This makes a direct comparison between the gas prescription of Agora and that of the other two simulations more challenging. However, there are things we can infer from differences in the calibration approaches used for each gas model. As in the previous models, BAHAMAS (and by extension Agora) also calibrate their simulation against X-ray gas mass fractions. However, in a departure from the previous models, BAHAMAS additionally calibrates simulation parameters to match measurements of the local Galaxy Stellar Mass Function (GSMF), based on observations from SDSS (\citealt{2013MNRAS.436..697B, 2009MNRAS.398.2177L}) and GAMA (\citealt{2012MNRAS.421..621B}). The full details of these calibrations can be found in \cite{2017MNRAS.465.2936M, Mead_2020} and we just highlight notable features here for comparison and completeness. 

Based on differences in calibration approaches, we can infer that Agora contains lower gas mass fractions and effectively higher levels of AGN feedback than either S10 or O22. In \cite{2017MNRAS.465.2936M}, X-ray gas fractions are mainly used to help specify the BAHAMAS AGN heating temperature, one of two parameters that determines the level of AGN feedback in the hydrodynamical simulation. While the GSMF is not found to scale sensitively with AGN heating temperature, the simulated gas-halo mass relation does. Adjustments of the AGN heating temperature in BAHAMAS results in three versions of the hydrodynamical simulation and the gas model of \cite{Mead_2020}. In pasting profiles, Agora opts for using the version of this model that best matches recent measurements of the tSZ angular power spectrum from SPT, $Planck$, and ACT (see Figure 15 in \citealt{omori2022}). This makes sense, given that the ultimate goal is to simulate an accurate Compton-$y$ map. However, the version Agora chose (BAHAMAS 8.0) has the highest AGN heating temperature (and AGN feedback), which results in a poorer match to the data of X-ray gas mass fractions (see Figure 4 in \citealt{2017MNRAS.465.2936M}) and is likely to cause discrepancies with S10 and O22. Previous works (e.g., \citealt{Shaw_2010}) have shown that increased feedback leads to a reduction in the overall amplitude of the tSZ power spectrum. We expect to observe this influence in the Agora tSZ power spectrum.

However, other parameter choices may compensate. For example, as discussed below, the cosmological parameters adopted for the MDPL2 simulation (in particular the higher value of $\Omega_\mathrm{m}$) would be expected to result in a higher tSZ power spectrum at fixed gas prescription compared to the other two simulations. This is because a higher value of $\Omega_\mathrm{m}$ results in a higher halo abundance (see, e.g., Equation 1 in \citealt{Bocquet_2015}). 

Additionally, in a departure from S10 and O22, the halo model of Agora includes tSZ contributions from a diffuse ejected gas component that resides outside of the halo virial radius. The presence of this ejected gas is directly related to feedback processes as well. Although, the ejected gas contributes only to the two-halo term, which is sub-dominant at $\ell = 3000$, it decreases the fractional amount of bound gas which has a significant effect on the one-halo term \citep{Mead_2020}. Both bound and ejected gas fractions in Agora are mass-dependent and described by Equations 25 and 26 in \citet{Mead_2020}, respectively. Notably, the fractional amount of ejected gas decreases with increasing halo mass, leading the fractional amount of bound gas to increase with halo mass. This effectively boosts the tSZ signal in higher-mass halos for Agora, in comparison to lower-mass halos. 

\begin{figure}
    \centering
    \includegraphics[width=3.3in]{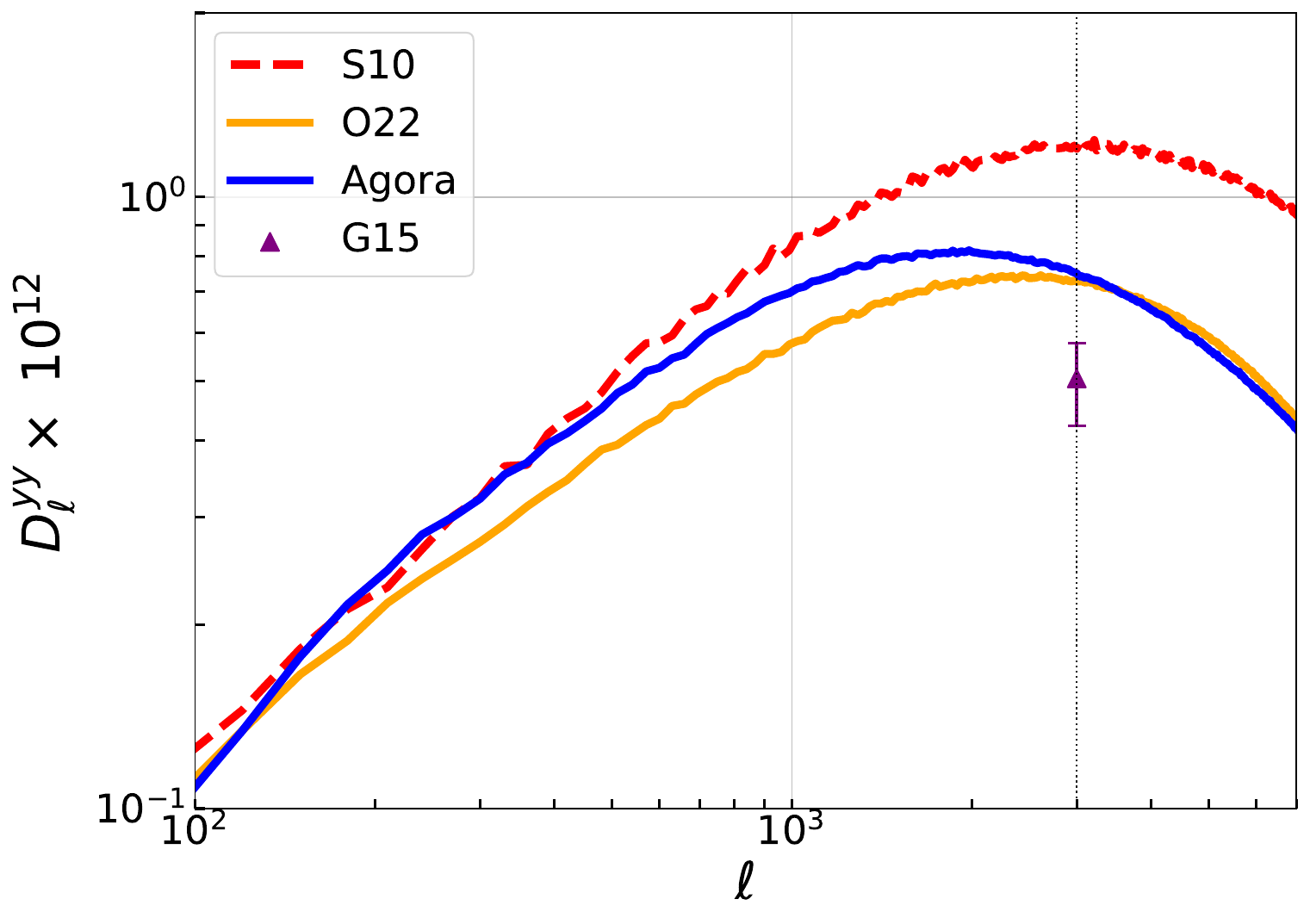}
    \caption{tSZ power spectra for simulation products and one data value used in this work. The curves represent full power spectra for each of the three simulated $y$-maps. The triangle marker, labelled "G15", denotes a full tSZ power spectrum amplitude derived from SPT-SZ data in \citealt{George_2015}. The dotted vertical dark gray line at $\ell=3000$ highlights the fiducial $\ell$ value used for subsequent analysis. Notably, all of the simulations predict a higher tSZ power spectrum amplitude than the data, with S10 having the highest amplitude at $\ell=3000$.}
    \label{fig:all_sim_tsz_ps}
\end{figure}

As in S10 and O22, Agora assumes a polytropic equation of state and hydrostatic equilibrium for their gas model, with an adiabatic index value of $\Gamma = 1.1966$. We refer the reader to \citet{Mead_2020} for a more detailed description of the other distinct gas model parameters and values employed. Agora predicts a tSZ power spectrum amplitude of $D^{yy}_{3000} = 7.41 \times 10^{-13}$ in units of Compton-$y$ at $\ell = 3000$. The full tSZ power spectrum for a wider range of $\ell$ values is shown in Figure \ref{fig:all_sim_tsz_ps}. \newline

The gas prescription for each simulation is only one component of the overall effect on the tSZ power spectrum. We must also consider cosmological effects related to the relative abundance of halos as a function of mass. Notably, there are some differences in the values of cosmological parameters adopted by each simulation, as can be seen in Table \ref{table:cosmoparam}, which may lead to differences in cluster abundance per mass and redshift. We consider both cosmological and gas model effects when anlyzing our baseline results, but also choose one metric for comparison that reduces the effects of cosmology and allows us to focus on gas model choices.

\begin{table*}
\centering
\begin{tabular}[t]{cccc}
\toprule 
Parameters & S10 & O22 & Description\\
\hline
$\Gamma$&1.2&1.2&Adiabatic Index\\
$\epsilon_{\rm DM}$&0.05&0&Energy transfer between dark matter and gas\\
$\epsilon_{*}$&$4\times10^{-6}$&$3.97\times10^{-6}$&Feedback from astrophysical sources\\
$f_{*}$&0.0164&0.026&Amplitude of stellar mass fraction scaling relation\\
$S_{*}$&0.26&0.12&Slope of stellar mass fraction\\
$\alpha_{\rm nt}$&0&0.18&Amplitude of non-thermal pressure\\
$\beta_{\rm nt}$&0&0.5&Redshift dependence of non-thermal pressure\\
$n_{\rm nt}$&0&0.8&Radial dependence of non-thermal pressure\\
\hline
\end{tabular}
\caption{Gas model parameters and values adopted by S10 and O22. Agora uses a different parameterization procedure, which is not as directly comparable to the other two simulations mentioned here, but is described in detail in \citealt{Mead_2020}.}
\label{table:gasparam}
\end{table*}

\begin{table*}
\centering
\begin{tabular}{l|l|l|l} \hline
\multicolumn{1}{l|}{}&\multicolumn{3}{l}{\textbf{Qualitative Treatment by Simulation}}\\\hline
\textbf{Astrophysical Quantity}&\textbf{S10}&\textbf{O22}&\textbf{Agora}\\\hline
Non-thermal Pressure&\begin{minipage}{3.5cm} ~\\ None. \\ \end{minipage} & \begin{minipage}{3.5cm} ~\\ Redshift- and radius- dependent.\\ \end{minipage} & \begin{minipage}{3.5cm} ~\\ None. \\
\end{minipage} \\\hline
AGN Feedback&\begin{minipage}{3.5cm} ~\\ More than O22. \\ \end{minipage} & \begin{minipage}{3.5cm} ~\\ Lowest amount. \\ \end{minipage} & \begin{minipage}{3.5cm} ~\\ Highest amount. \\
\end{minipage} \\\hline
Dark Matter Energy Transfer&\begin{minipage}{3.5cm} ~\\ Has similar effects as AGN feedback. \\More than O22. \\ \end{minipage} & \begin{minipage}{3.5cm} ~\\ None \\ (i.e., parameter value set to zero). \\ \end{minipage} & \begin{minipage}{3.5cm} ~\\ None, explicitly \\ (i.e., no parameter). \\
\end{minipage} \\\hline
Bound-gas fraction&\begin{minipage}{3.5cm} ~\\ All gas is bound. \\Always equal to 1. \\ \end{minipage} & \begin{minipage}{3.5cm} ~\\ Always equal to 1.\\ \end{minipage} & \begin{minipage}{3.5cm} ~\\ Strong scaling \\ with mass. \\ 
\end{minipage} \\\hline
Stellar-mass fraction&\begin{minipage}{3.5cm} ~\\ Mass-dependent, with \\ shallower slope than\\ O22. \\ \end{minipage} & \begin{minipage}{3.5cm} ~\\ Highest overall amplitude with steepest slope and mass-dependence.\\ \end{minipage} & \begin{minipage}{3.5cm} ~\\ Lowest amplitude with weak redshift- and mass- dependence.\\
\end{minipage} \\\hline
\end{tabular}
\caption{Qualitative summary of relevant astrophysical quantities from each simulation. More details in the text and in the associated papers of each product.} \label{astro_quantities_description}
\end{table*}

\section{Data}
\label{sec:data}
The SPT-SZ survey is a $\sim$2500 ${\rm deg}^{2}$ survey of the southern sky with three observation bands centered at frequencies of roughly 95, 150, and 220 GHz \citep{Carlstrom_2011}. In this work we make use of Compton-$y$ maps presented in \citet{Bleem_2022}, specifically using the ${\rm half}1$ and ${\rm half}2$ ``minimum-variance'' maps.\footnote{While these maps are referred to as ``minimum-variance maps'' in \citet{Bleem_2022}, they have in fact been weighted slightly sub-optimally in order to reduce CIB contamination. We discuss this point in more detail in Section~\ref{sec:robustness_tests}.} These Compton-$y$ maps have an angular resolution of 1.25' and are constructed from both SPT-SZ and $Planck\ 2015$ data \citep{2016}. By combining data from both surveys, the noise level of the resulting Compton-$y$ is minimized over a wide range of angular scales. The ${\rm half}1$ and ${\rm half}2$ Compton-$y$ maps are each created using half of the full data set. The advantage of using half maps over the full $y$ map is that taking their cross-spectra nulls any uncorrelated instrumental noise. In our analysis, we mask the top 5$\%$ of galactic dust regions, determined using a CMB-subtracted $Planck \ 545$ GHz foreground map presented in \citet{2016} as a template.  Furthermore, we also mask bright emissive point sources detected above $6$ mJy (at 150 GHz). We refer the reader to \cite{Bleem_2022} for more details on both dust and point-source masks. Note that throughout this paper we use the terms "masked" and "unmasked" in reference to galaxy clusters only; dust and point-source masking is assumed throughout for the data. The cluster catalog we use to select what clusters to mask as a function of mass and redshift in this $y$ map is also derived from SPT-SZ data. This catalog is described in \citet{Bleem_2015, Bocquet_2019} and includes spectroscopic redshifts and $M_{500}$ estimates (based on an SZ observable-mass scaling relation) for 516 identified clusters.

\begin{figure}
    \centering
    \includegraphics[width=3.33in]{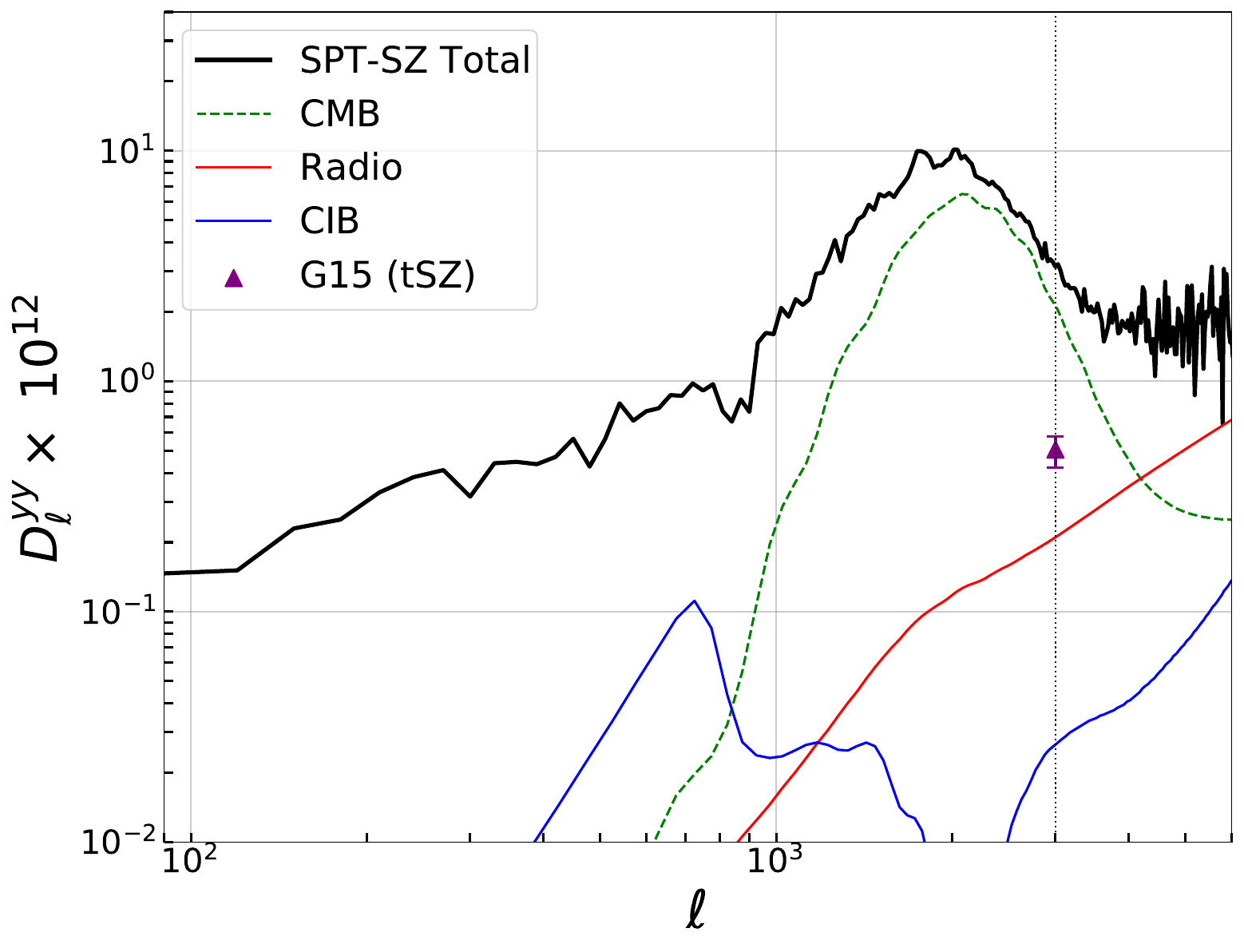}
    \caption{Data $y$ map power spectrum along with the spectra of model contaminants. The black line denotes the cross-power-spectrum of the ${\rm half}1$ and ${\rm half}2$ minimum-variance Compton-$y$ maps from \citet{Bleem_2022}. Model contaminant curves are derived in the same manner as used for Figure 7 of \citet{Bleem_2022}, with different frequency weights applied to match the minimum-variance $y$ map used here. As in Figure \ref{fig:all_sim_tsz_ps}, the triangle marker denotes a full tSZ power spectrum amplitude, derived from SPT-SZ data in G15, and which we adopt for some subsequent calculations. The dotted vertical dark gray line at $\ell = 3000$ represents the fiducial $\ell$ value used for analysis in this work.}
    \label{fig:y_map_halves_PS}
\end{figure}

As in the previous section, we use as a reference the tSZ amplitude measured in \citet{George_2015}, which we henceforth refer to as G15.\footnote{We use measurements from G15 over more recent measurements from \citet{2021ApJ...908..199R} mainly due to the fact that our data $y$ map was created using the same data set that G15 uses. Furthermore, G15 describes the results of a cluster masking test, which we use in a robustness test described in Section \ref{G15_comparison}} This measurement has a value of $D^{yy}_{3000}=(5.05 \pm 0.78) \times 10^{-13}$ in units of Compton-$y$ and is denoted by a purple triangle marker in Figures \ref{fig:all_sim_tsz_ps} and \ref{fig:y_map_halves_PS}. To derive this value, \citet{George_2015} adopts a baseline model from \citet{Shaw_2010} as a template in Markov chain Monte Carlo (MCMC) chains in order to constrain the tSZ amplitude from SPT-SZ data. Additional constraints are applied using the 800 ${\rm deg}^{2}$ SPT bispectrum measurement described in \citet{Crawford_2014}.

In Figure \ref{fig:y_map_halves_PS}, we show the cross-power-spectrum between the ${\rm half}1$ and ${\rm half}2$ SPT-SZ minimum-variance Compton-$y$ maps as a black line. The dotted vertical dark gray line at $\ell = 3000$ represents the fiducial $\ell$ value used for comparison in our baseline results. We note that the amplitude of this power spectrum, at $\ell=3000$, is almost an order of magnitude larger than that of the G15 data and the three simulations considered in Figure \ref{fig:all_sim_tsz_ps}. This is primarily indicative of other microwave sources which
have not been completely selected out of the Compton-$y$ maps and which effectively act as sources of contamination. These include signals from radio galaxies, dusty galaxies that make up the cosmic infrared background (CIB), and the primary CMB.

\section{Methods}
\label{sec:methods}

In our analysis we mask galaxy clusters in several mass and redshift bins and use these to compute three main quantities for comparison. To this end, we make use of Python \texttt{healpy} utilities \citep{Zonca2019}, as well as halo masses, redshifts, and positions provided in the associated cluster catalogs of each simulation and data product. We take $M_{500}$ to be our default halo mass, where this quantity is defined as
\begin{equation}
    M_{500} = 500\rho_{\rm crit}(z) \frac{4\pi}{3} R_{500}^{3},
\label{eq:M500c}
\end{equation}
with $\rho_{\rm crit}$ being the critical matter density of the universe and $R_{500}$ being the characteristic radius at which this equation is satisfied (i.e., the radius at which the cluster's spherical overdensity is equal to 500 times the critical density of the universe).  

For simplicity, we use binary tophat masks (not apodized) throughout. These also have the advantage of reducing any potential masking of residual tSZ signal from filamentary structure in data. In Section \ref{sec:robustness_tests} we test the effects of varying this procedure and find that our baseline method is robust enough for the purposes of our analysis here. Individual masks are tailored to depend on each cluster's characteristic $\theta_{500}$ value, which can be defined as
\begin{equation}
    \theta_{500} = \frac{R_{500}}{D_{\rm A}}
\label{eq:5}
\end{equation}
where $D_{\rm A}$ is the angular diameter distance, which depends on redshift and cosmology. We specifically use the COLOSSUS toolkit \citep{Diemer_2018} to compute $D_{\rm A}$ and $R_{500}$, using catalog $M_{500}$ and $z$ values, and thus determine $\theta_{500}$ for each halo.  We use $\theta_{500}$ in units of arcminutes ($'$) and apply the following masking routine: 
for $\theta_{500} \geq 2'$, we use a mask radius of $2 \times \theta_{500}$, whereas for $\theta_{500} < 2'$, we use a mask radius of $4'$. This routine is motivated by our understanding of the radial extent of most of the tSZ signal in data and simulations. 

When masking out galaxy clusters by mass and redshift in all Compton-$y$ maps, a mask correction factor is applied to correct for holes in the map area and refer cut-sky estimates to the full sky. In the case of the data, a beam correction factor is also applied to power spectra via healpy utilities in order to properly compare to simulations. Similarly, healpy tools correct for differences in the pixel window function of both data and simulations.

The first quantity used to measure the effects of masking clusters is simply the difference between the masked and unmasked\footnote{Once again, "masked" and "unmasked" in this case refers only to galaxy clusters. Dust and radio point-sources are always masked for the data.} tSZ power spectra at $\ell=3000$, which we describe with 
\begin{equation}
    \Delta D_{3000}^{yy} (M_{500}) = D_{3000,{\rm unmasked}}^{yy}-D_{3000,{\rm masked}}^{yy}(M_{500})
\label{eq:D3000}
\end{equation}
This metric allows us to specifically probe and compare the contributions of high-mass clusters ($M_{500} > 1 \times 10^{14}\ M_\odot\ h^{-1}$) to the overall tSZ power spectrum amplitude in each $y$ map. Beyond individual cluster gas physics, $\Delta D_{3000}^{yy}$ measurements are also dependent on the cluster count in each of our mass and redshift bins. These cluster counts are reflective of the cosmology and halo-abundance in each data and simulation product. 

In order to minimize cosmological effects, and focus on studying the effects of gas physics, we additionally normalize $\Delta D_{3000}^{yy}$ relative to one galaxy cluster in each bin. We refer to this quantity as $\Delta D_{3000}^{yy} \ {\rm per \ cluster}$ and use it as a second metric for comparison. To measure this, we divide $\Delta D_{3000}^{yy}$ values by the cluster count in each mass and redshift bin, normalized relative to a full sky area, such that
\begin{equation}
    \Delta D_{3000}^{yy} \ {\rm per \ cluster} \ (M_{500}) = \frac{\Delta D_{3000}^{yy} (M_{500})}{N_{\rm clusters, \ full-sky}}
\label{eq:D3000_per_cluster}
\end{equation}

In order to infer the contributions to the total power from low-mass clusters, we incorporate full tSZ power spectrum amplitude measurements into our third, and final, metric for comparison. We specifically divide $\Delta D_{3000}^{yy}$ values by the full tSZ power spectrum amplitude of each respective product at $\ell = 3000$. We refer to this quantity as the fractional power contribution at $\ell=3000$ ($f_{3000}$) and define it as:
\begin{equation}
    f_{3000} (M_{500}) = \frac{\Delta D_{3000}^{yy} (M_{500})}{D_{3000}^{yy,tSZ}}
\label{eq:f3000}
\end{equation}
For simulated $y$ maps, $D_{3000}^{yy,tSZ}$ in Equation \ref{eq:f3000} is equivalent to $D_{3000,{\rm masked}}^{yy}$ in Equation \ref{eq:D3000}. However, since our data contains additional sources of microwave contamination (see Figure \ref{fig:y_map_halves_PS}), we require a different $D_{3000}^{yy,tSZ}$ value that better approximates the true total amplitude of the tSZ power spectrum for our treatment of the data. In our analysis we reference the aforementioned tSZ power spectrum amplitude measurement from G15, which has a value of $D_{3000}^{yy,tSZ}=(5.05 \pm 0.78) \times 10^{-13}$ in units of Compton-$y$. To be clear, we still take the difference between masked and unmasked power spectra for our data $y$ maps, but we divide this quantity by the full amplitude from G15 to obtain $f_{3000}$ values as a function of mass. 

As previously mentioned, we compute $\Delta D_{3000}^{yy}$, $\Delta D_{3000}^{yy} \ {\rm per \ cluster}$, and $f_{3000}$ for data and simulations in several mass and redshift bins. Since cluster abundance decreases exponentially with halo mass \citep{Tinker_2008}, we use logarithmically spaced mass bins. Specifically, we initially create nine mass bins in the range $1 \times 10^{14}\ M_\odot\ h^{-1} < M_{500} < 1 \times 10^{15}\ M_\odot\ h^{-1}$. We then manually widen our last bin up to $2 \times 10^{15}\ M_\odot\ h^{-1}$ in order to accommodate all high-mass clusters. For full-redshift values we use the range $0.25<z<1.5$. Our lower redshift boundary is mostly motivated by the SPT-SZ cluster selection function (see, e.g., \citealt{Bleem_2015}). The upper limit is motivated by the maximum redshift of massive clusters in the Agora catalog at the initial time of analysis.\footnote{The Agora catalog has since been updated to include higher redshift clusters.} For our redshift-binned values we split up our cluster sample between two $z$-bins defined by: $0.25<z<0.59$ and $0.59<z<1.5$. These were chosen so as to divide our cluster counts as evenly as possible and thus reduce uncertainties on our y-axis values. Again, we apply mask-correction factors throughout, so as to calculate these values in the full-sky regime.

In addition to our mass and redshift bins we use $\ell$ bins to compute mean values for each of our main metrics for comparison. For each mass and redshift bin, we define five evenly-spaced $\ell$ bins in the range $2950 \leq \ell \leq 3050$. A bin size of $\Delta \ell = 20$ is chosen due to the fact that, because of the size of its observation region, the SPT-SZ survey has a resolution limit of $\Delta \ell \sim 10$. We measure our metric value in each of the five $\ell$ bins and then take the average of all five to obtain a mean metric value for comparison.

We estimate the uncertainty on $\Delta D_{3000}^{yy}$ through:
\begin{equation}
    \sigma_{\Delta D_{3000}^{yy}}^{2} = \frac{{\rm Var}(\Delta D^{yy}_{\ell})}{N_{\ell,{\rm bins}}} + \frac{(\Delta D_{3000}^{yy})^{2}}{N_{\rm clusters}},
\label{eq:error2}
\end{equation}
where ${\rm Var}(\Delta D^{yy}_{\ell})$ is the variance in individual $\ell$ bin values of $\Delta D^{yy}_{\ell}$, $N_{\ell,{\rm bins}}$ is the number of $\ell$ bins (i.e., five), and $N_{\rm clusters}$ is the cluster count in each mass and redshift bin. 
The first term in Equation~\ref{eq:error2} is our estimate of the variance on the mean value across the $\ell$ bins, and the second term is from Poisson statistics. %$N_{\ell,{\rm bins}} = 5$ and is the aforementioned number of $\ell$ bins used to calculate each mean $\Delta D^{yy}_{3000}$ value, while 
The uncertainty on the data in most mass bins is dominated by the Poisson term.
%in Equation \ref{eq:error2}. 

To approximate the error on $\Delta D_{3000}^{yy} {\rm \ per \ cluster}$, we additionally divide $\sigma_{\Delta D_{3000}^{yy}}^{2}$ by the full-sky cluster count in each mass and redshift bin. Similarly, the uncertainty on $f_{3000}$ is $\sigma_{\Delta D_{3000}^{yy}}^{2}$ divided by $D_{3000,{\rm unmasked}}^{yy}$. As in Equation \ref{eq:f3000}, we adopt a $D_{3000,{\rm unmasked}}^{yy}$ value from G15 for our treatment of the data.

Additionally, for the data, we approximate uncertainties on our mean cluster mass values in each bin by calculating the propagation of error due to uncertainties on the SPT-SZ mass-significance relation. This relation is described by Equation 1 in \citet{Bocquet_2019}, for which a fiducial $\nu \Lambda$CDM cosmology is employed, with parameter values and uncertainties described in column 2 of Table 3 of the same paper. For clarity, we show this equation here:
\begin{equation}
%\begin{aligned}
    \langle {\rm ln} \zeta \rangle = {\rm ln}A_{\rm SZ} \ + \ B_{\rm SZ} {\rm ln}\Big(\frac{M_{500} h_{70}}{4.3 \times 10^{14} M_\odot\ }\Big) 
    \\ + \ C_{\rm SZ}{\rm ln}\Big(\frac{E(z)}{E(0.6)}\Big)
\label{eq:mass-relation}
%\end{aligned}
\end{equation}
To obtain uncertainty measurements from this we first compute mean significance values for each of our mass bins. We then invert this relation to solve for $M_{500}$, as a function of $\zeta$, and propagate the error on the parameters $A_{\rm SZ}$, $B_{\rm SZ}$, and $C_{\rm SZ}$ to approximate the error on $M_{500}$. A fiducial redshift value of $z = 0.7$ is used throughout. This treatment effectively measures the mean correlated uncertainty value for each mass bin. We assume that there are enough clusters in each mass bin so as to make the uncorrelated uncertainty negligible.

\section{Baseline Results}
\label{sec:baseline_results}

In this section we present baseline results for our three main metrics for comparison between data and simulations. 
In generating baseline values we convert all halo masses into units of $\ M_\odot\ h^{-1}$ for easier comparison. For consistency, we assume $h=0.7$ throughout. 

We note that the $>95 \%$ cluster mass-completeness limit for the SPT-SZ survey, and its associated cluster catalog, is cited as $M_{500} \sim  7 \times 10^{14}\ M_\odot\ h^{-1}_{70}$ \citep{Bleem_2015}, corresponding to a value of $M_{500} \sim  5 \times 10^{14}\ M_\odot\ h^{-1}$. This means that any sample of clusters identified below this mass is not a complete sample, and thus the total tSZ power contribution of all clusters below this mass cannot be directly estimated without implementing some kind of completeness correction. This becomes increasingly difficult to do at lower mass values, due to uncertainties in our completeness fractions, and so we only apply this treatment to one data point near the completeness limit in each relevant baseline figure.  

\subsection{$\Delta D_{3000}^{yy}$ }

In the top panel of Figure \ref{fig:diff3000_differential_final} we show our baseline full-redshift $\Delta D_{3000}^{yy}$ calculations, for SPT-SZ data and all three simulations considered. The shaded region in Figure \ref{fig:diff3000_differential_final} represents the mass incompleteness regime of clusters in the SPT-SZ survey of \citet{Bleem_2015}. We include three SPT-SZ data points in the non-shaded $>95 \% $ complete mass regime, and one point in the shaded region (near-completeness). We also apply a completeness correction to this last point. We do this by dividing the incomplete $\Delta D_{3000}^{yy}$ value by the estimated completeness fraction for $y$ power spectrum measurements in that bin. We note that this completeness fraction is not identical to the completeness for cluster-finding. To estimate the completeness for $y$ power spectrum measurements, we make use of existing SPT-SZ mock-observations of $y$ maps from the MAGNETICUM simulation \citep{Dolag_2016}. We compute power spectra from MAGNETICUM $y$ maps with and without masking clusters, but in any one mass bin we only mask clusters that exceed the \citet{Bleem_2015} detection threshold in mock-observed maps.
 The bottom panel of Figure \ref{fig:diff3000_differential_final} is a histogram of cluster counts from each simulation and data product, normalized to the SPT-SZ survey area. 

In Figure \ref{fig:diff3000_differential_final_zbinned} we show our baseline redshift-binned $\Delta D_{3000}^{yy}$ calculations and cluster counts. This is similar to Figure \ref{fig:diff3000_differential_final}, except our cluster sample is broken up into our two redshift bins, defined as: $0.25<z<0.59$ and $0.59<z<1.5$. Completeness corrections are applied as previously outlined.

\subsection{$\Delta D_{3000}^{yy} {\rm \ per \ cluster}$} 

In Figures \ref{fig:diff3000_per_cluster} and \ref{fig:diff3000_per_cluster_zbinned}, we present baseline $\Delta D_{3000}^{yy} \ {\rm per \ cluster}$ values. Figure \ref{fig:diff3000_per_cluster} shows these values for our full-redshift range, while Figure \ref{fig:diff3000_per_cluster_zbinned} displays these values for our two redshift bins. This metric represents the contribution to the tSZ power spectrum from an average cluster in each mass bin. To calculate this, we divide $\Delta D_{3000}^{yy}$ values in a given bin by the cluster count in that bin, normalized relative to a full sky area. This assumes that all clusters in a mass and redshift bin contribute proportionally the same to the tSZ power spectrum at $\ell = 3000$. By using the $\Delta D_{3000}^{yy}$ per cluster as a basis of comparison, we can isolate the effects of different gas prescriptions from factors such as halo abundance and cosmology.

\subsection{$f_{3000}$}

In Figures \ref{fig:f3000_differential_final} and \ref{fig:f3000_differential_final_zbinned} we show baseline $f_{3000}$ values for our full redshift range and our two redshift bins, respectively. This metric is described by Equation \ref{eq:f3000}, and amounts to dividing the $\Delta D_{3000}^{yy}$ values for each simulation or data product by the respective full tSZ power spectrum amplitude at $\ell = 3000$ measured from each product. All masking and binning routines remain the same. This metric allows us to incorporate the tSZ power contributions of low-mass clusters into our measurements and comparisons. Completeness corrections are applied to one of our data values as outlined before. As a reminder, in our analysis, we adopt a full tSZ power spectrum amplitude value from G15 for the SPT-SZ data. This choice is appropriate given that the G15 value was derived using the same data used to produce our baseline $y$ map.

\begin{figure*}
    \centering
    \includegraphics[width=5in]{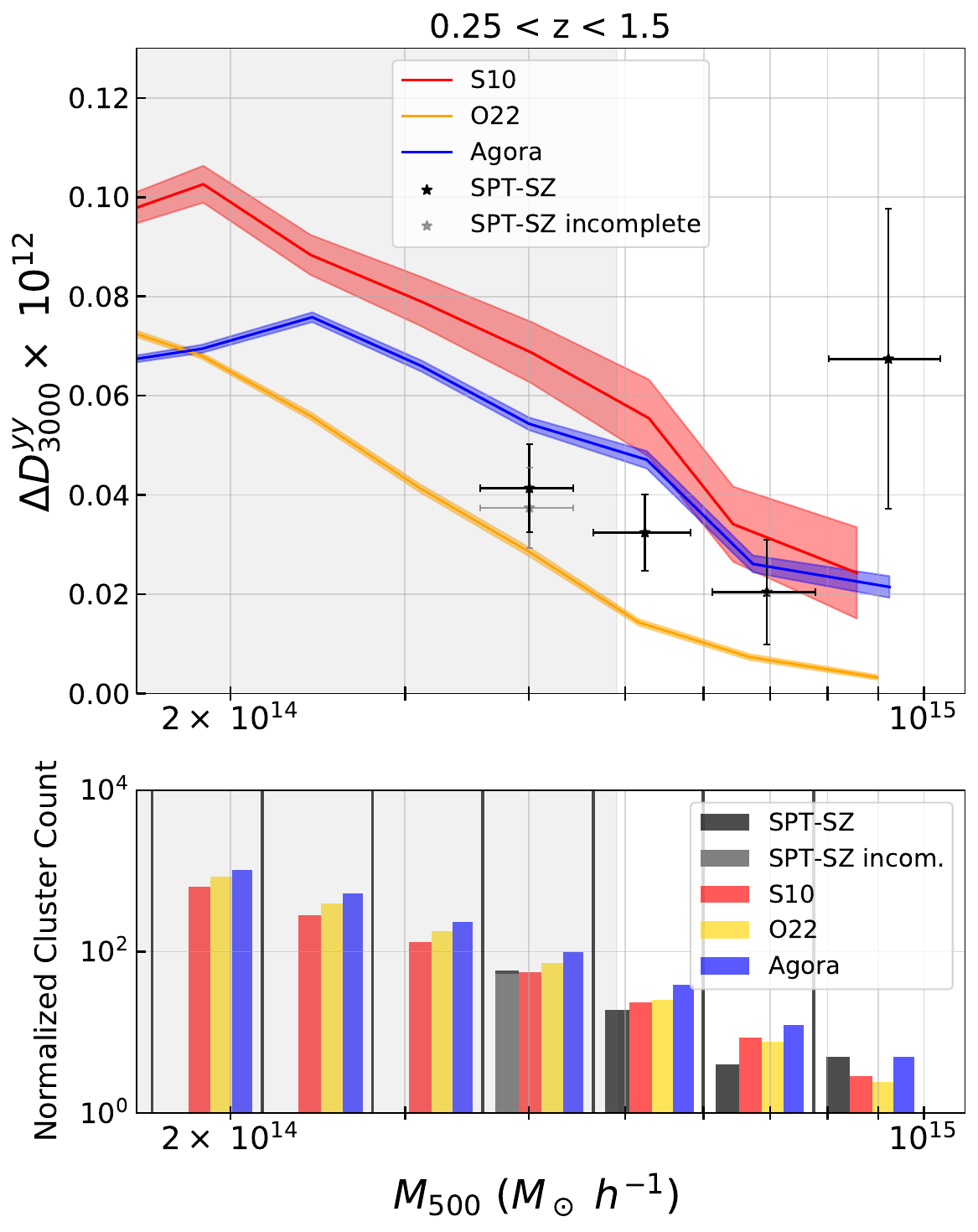}
    \caption{\textit{Top panel}: Full-redshift values of the difference between masked and unmasked power spectra, $\Delta D_{3000}^{yy}$, normalized relative to the full sky. The gray shaded region represents the mass-incomplete regime of the SPT-SZ cluster catalog of \citealt{Bleem_2015, Bocquet_2019}. The redshift boundary implemented is specified at the top of the figure. A completeness correction is applied to the lowest-mass SPT-SZ data point and denoted by a black star marker. The original data point, without this correction, is denoted by a gray star marker. The observed trends and outcomes presented here can be ascribed to a combination of cosmological and gas model choices, as detailed in the text. \textit{Bottom panel}: A histogram of cluster counts in each mass bin, normalized relative to the area of the SPT-SZ observation region. The solid black vertical lines represent the boundaries of our last seven mass bins; the first three mass bins are omitted for simplicity and since they are less relevant to the data.  We still include three bins without data values for a comparison between simulation models at lower masses. The highest mass bin is manually widened to include all high-mass clusters. Notably, Agora consistently predicts the highest cluster abundance here, which, along with the other relative cluster counts, helps explain some of the results in the top panel.} 
    \label{fig:diff3000_differential_final}
\end{figure*}

\begin{figure*}
    \centering
    \includegraphics[width=7.5in]{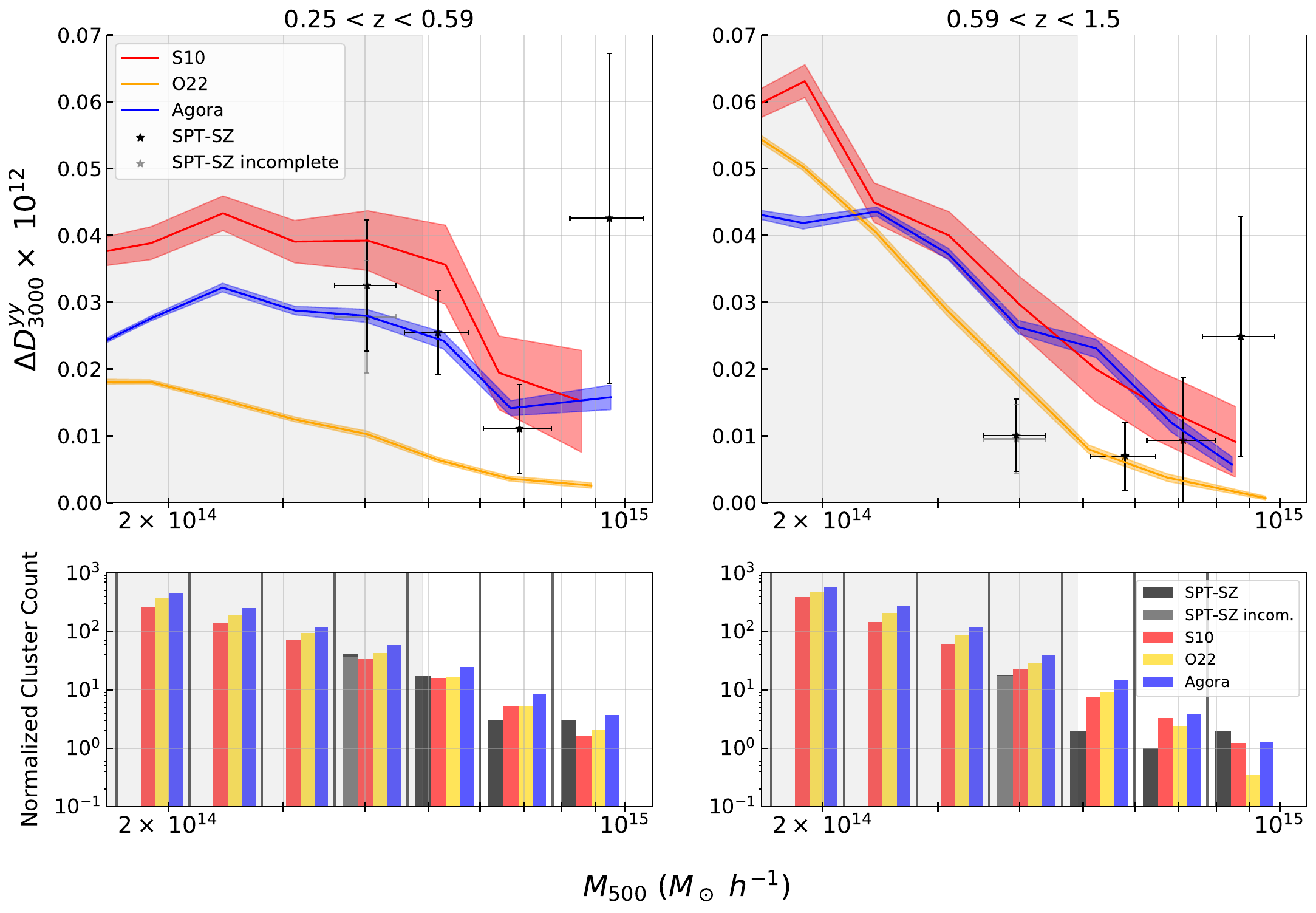}
    \caption{Redshift-binned $\Delta D_{3000}^{yy}$ values. Similar to Figure \ref{fig:diff3000_differential_final}, but with additional redshift-binning of clusters. The specific boundaries of our two redshift bins are specified in the title of this figure and are defined as: $0.25 < z < 0.59$ and $0.59 < z < 1.5$. These boundaries are chosen so as to split up our cluster count as evenly as possible. Shaded regions denote the mass-incomplete regime of the data. Here O22 exhibits the lowest $\Delta D_{3000}^{yy}$ amplitude overall, yet notably provides a closer match to the data and other simulations at higher redshift. On the other hand, Agora and S10 display higher amplitudes and provide a better fit to the data at lower redshift. Again, these trends can be attributed to a combination of cosmological and gas model choices, as discussed in the text.} 
    \label{fig:diff3000_differential_final_zbinned}
\end{figure*}

\begin{figure*}
    \centering
    \includegraphics[width=4.9in]{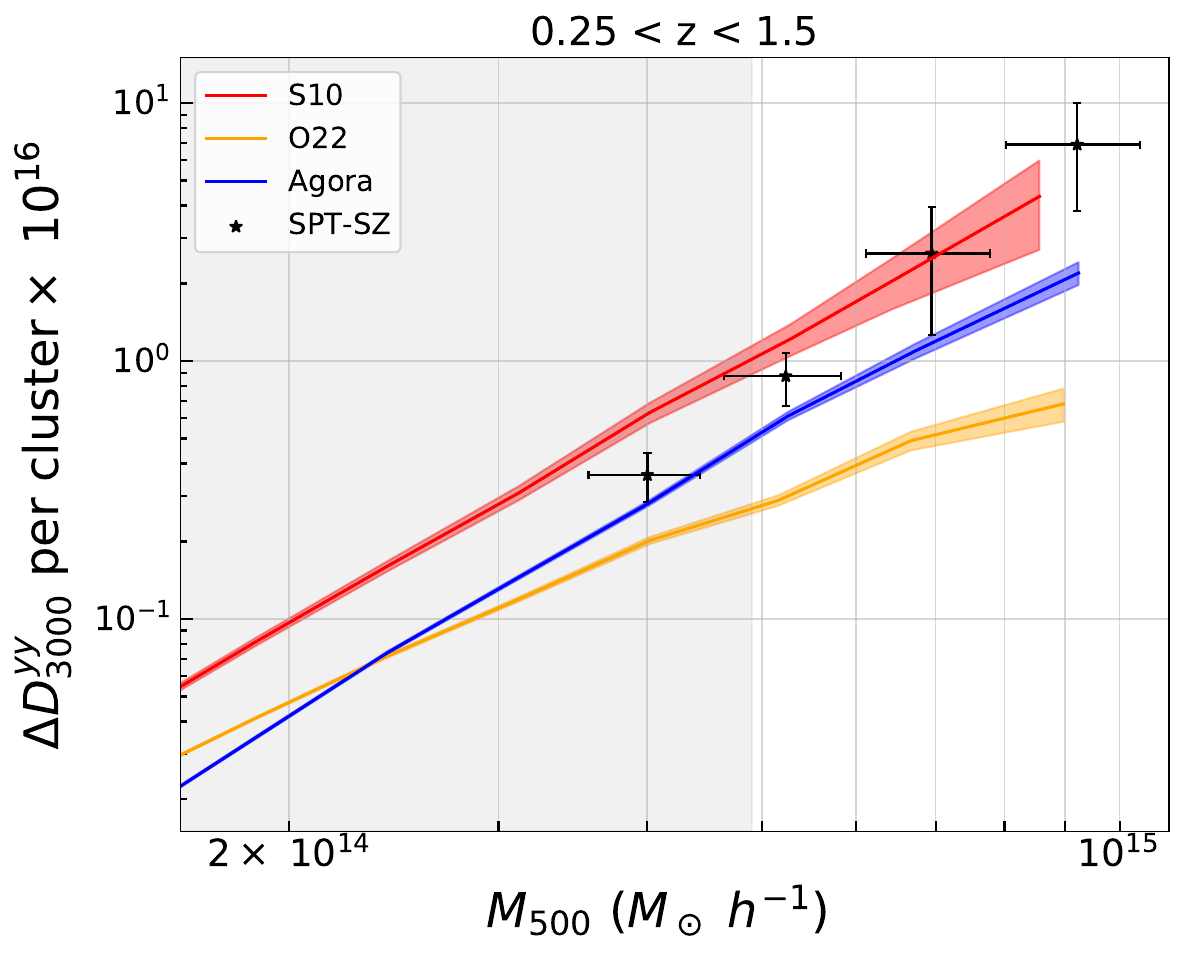}
    \caption{Full-redshift values of $\Delta D_{3000}^{yy} \ {\rm per \ cluster}$. Similar to the top panel of Figure \ref{fig:diff3000_differential_final}, but normalized relative to one cluster from each simulation and data product. This measurement is achieved by dividing $\Delta D_{3000}^{yy}$ values from Figure \ref{fig:diff3000_differential_final} by full-sky cluster counts for each mass bin. This metric facilitates a more direct comparison of the effects of gas model choices. Notably, S10 provides a closer match to the data at higher masses, while Agora provides a better match at lower masses. Conversely, O22 provides the worst fit overall. These findings likely arise due to different degrees of feedback, among other distinct gas model features, within each simulation, as elaborated upon in the text.} 
 \label{fig:diff3000_per_cluster}
\end{figure*}

\begin{figure*}
    \centering
    \includegraphics[width=7.1in]{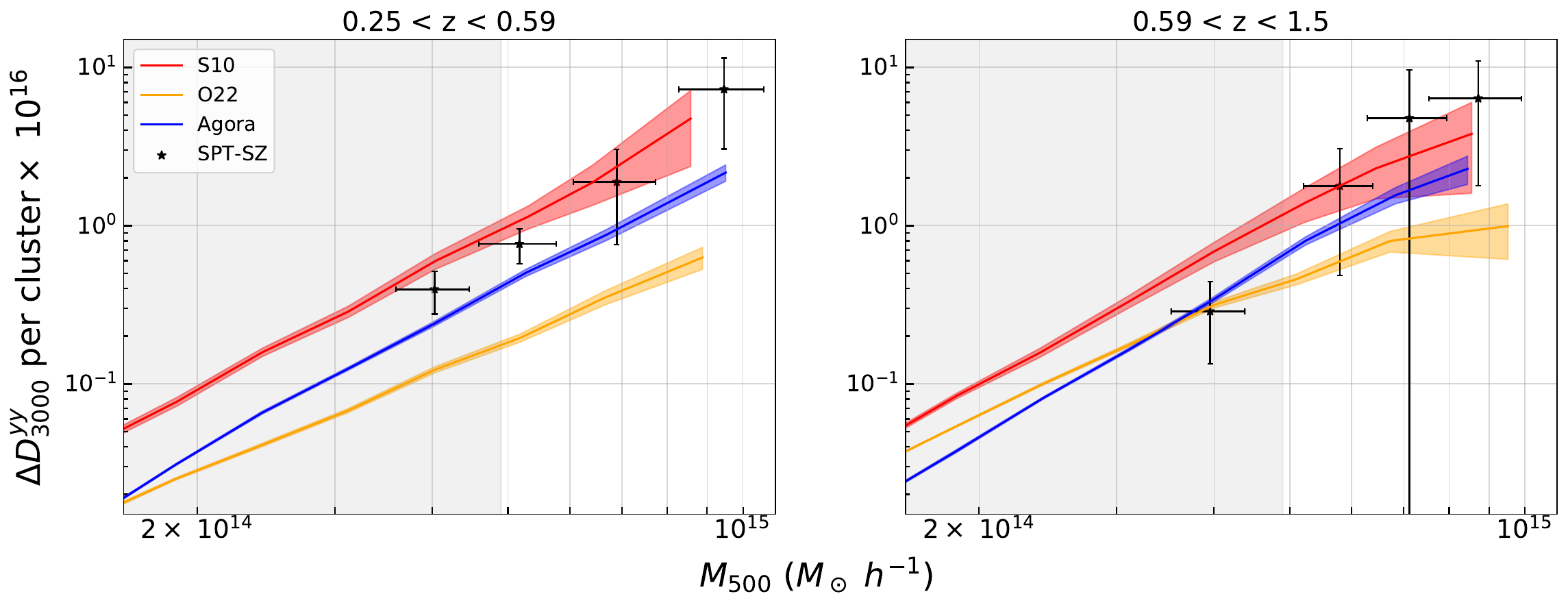}
    \caption{Redshift-binned values of $\Delta D_{3000}^{yy} \ {\rm per \ cluster}$. Similar to the top panel of Figure \ref{fig:diff3000_differential_final_zbinned}, but normalized relative to one cluster from each simulation and data product. Notably, the higher uncertainty associated with high redshift data points results in rough agreement between the data and all simulations at high redshift. Nevertheless, S10 and Agora exhibit better agreement with the data at lower redshift, likely owing to reasons similar to those outlined for the full-redshift values in the text.} 
 \label{fig:diff3000_per_cluster_zbinned}
\end{figure*}

\begin{figure*}
    \centering
    \includegraphics[width=5in]{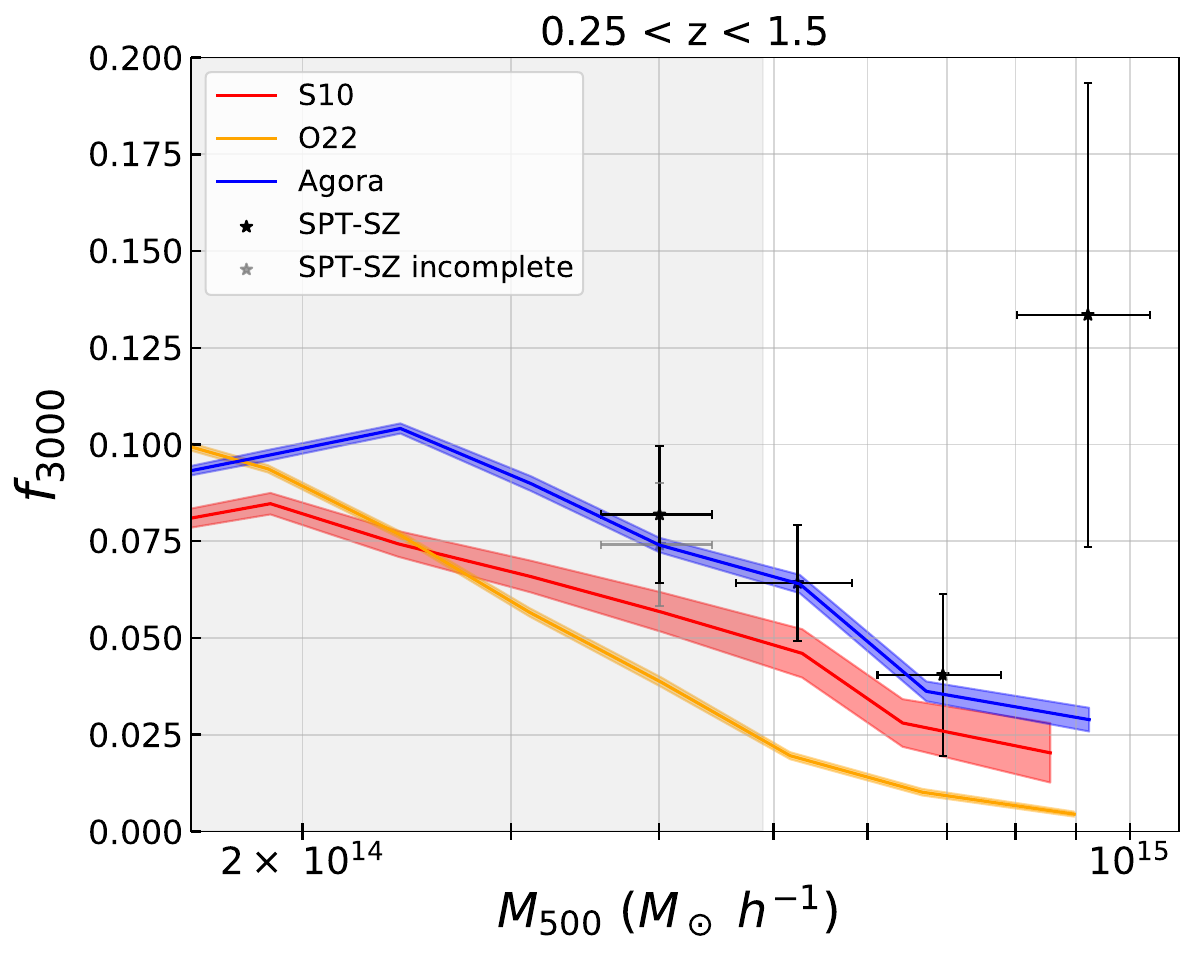}
    \caption{The fractional power contribution ($f_{3000}$) from clusters as a function of halo mass in our full-redshift bin. Similar to the top panel of Figure \ref{fig:diff3000_differential_final} but normalized by the full tSZ power spectrum amplitude of each product as described in Equation \ref{eq:f3000} and seen in Figure \ref{fig:all_sim_tsz_ps}. This metric provides insight into the relative influence of both low- and high-mass clusters on the tSZ power spectrum. Notably, Agora's mass-dependent treatment of their bound gas fraction contributes to a higher proportion of tSZ signal at high masses, resulting in a closer match to the data.}
\label{fig:f3000_differential_final}
\end{figure*}

\begin{figure*}
    \centering
    \includegraphics[width=7.1in]{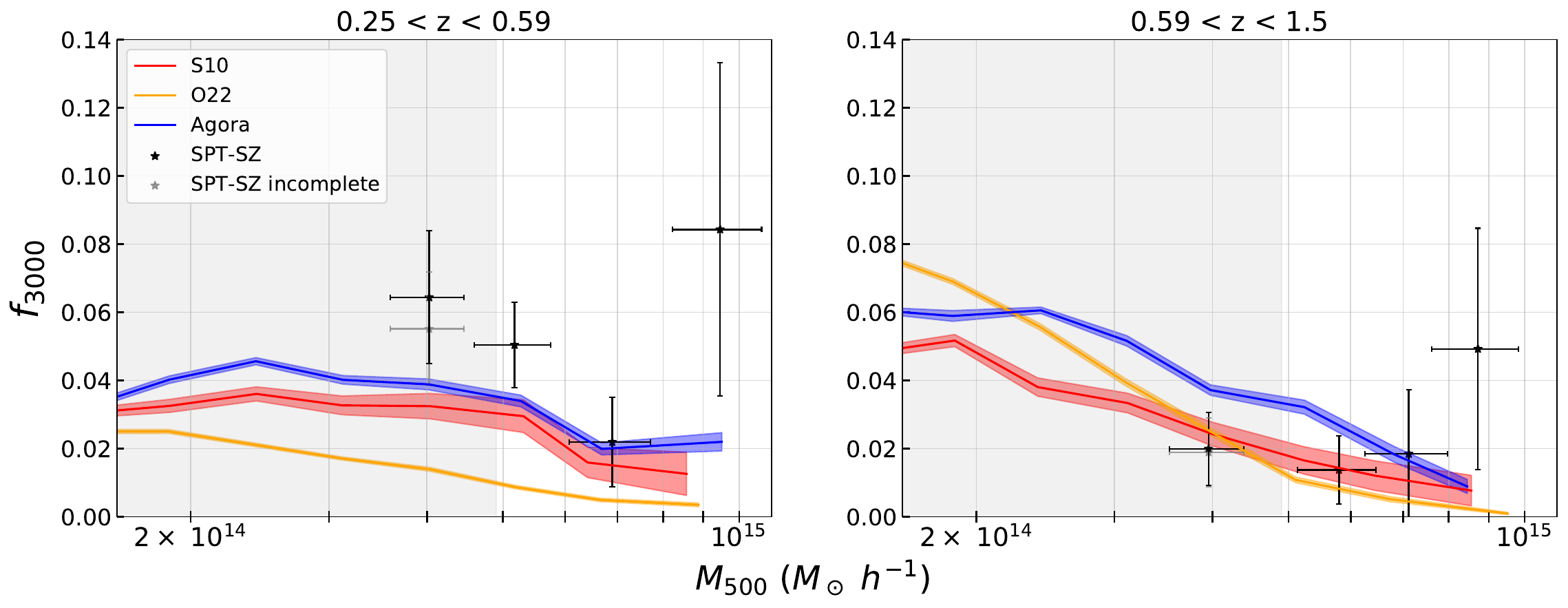}
    \caption{Redshift-binned $f_{3000}$ values. Similar to Figure \ref{fig:diff3000_differential_final_zbinned}, but with the inclusion of full tSZ power spectrum amplitudes as described in Equation \ref{eq:f3000} and seen in Figure \ref{fig:all_sim_tsz_ps}. Remarkably, higher redshift simulation values provide a better match to the data than lower redshift values.} 
    \label{fig:f3000_differential_final_zbinned}
\end{figure*}

\section{Discussion}
\label{sec:summary_and_discussion}

In this section, we provide a summary and discussion of our baseline results. We analyze the findings for each of our metrics individually and then draw general conclusions from all three in Section \ref{sec:conclusion}. To make more quantitative comparisons, we introduce a $\chi^{2}$-like statistic for each full-redshift metric and simulation value. This statistic takes into account the uncertainties of both the data and simulations and is defined as follows:
\begin{equation}
    \chi^{2}{\rm \mhyphen like} = \sum \frac{({\rm value}_{\rm data} - {\rm value}_{\rm sim})^2}{\sigma_{\rm data}^2 + \sigma_{\rm sim}^2} 
\label{eq:Xi-like}
\end{equation}
Here, ${\rm value}_{\rm data}$ represents any one metric value for the data, ${\rm value}_{\rm sim}$ represents the corresponding metric value for a specific simulation, and $\sigma_{\rm data}^2$ and $\sigma_{\rm sim}^2$ represent the squares of their respective uncertainty values.

Table \ref{agreements_between_sims_&_data} serves as an organizational tool, providing a qualitative summary of the agreement regimes between the data and simulations, along with potential interpretations.

\subsection{$\Delta D_{3000}^{yy}$}

For the full-redshift $\Delta D_{3000}^{yy}$ values (Figure \ref{fig:diff3000_differential_final}), we observe that S10 generally has the highest amplitude, followed by Agora and then O22. The higher overall amplitude of S10 can be attributed to the absence of non-thermal pressure support in its gas model and a lower amount of feedback compared to Agora. Agora, on the other hand, has a higher amplitude than O22 due to both gas model choices (discussed more in-depth in the next subsection) and a higher halo abundance per mass bin. It is worth noting that Agora has the highest halo abundance in all mass bins, except for the last bin where it is comparable to the data. This is reflective of MDPL2’s choice of cosmological parameters, in particular the higher value of $\Omega_{\rm m}$ used compared to the other simulations (see Section \ref{subsec:Agora_gas}). In terms of agreement with the data, none of the simulation values for full-redshift $\Delta D_{3000}^{yy}$ strongly align with the data. However, the data generally falls between the values of Agora (upper bound) and O22 (lower bound), with slightly better agreement observed with Agora. The calculated $\chi^{2}$-like statistic value for Agora is $8.19$, compared to $13.76$ for O22 and $13.82$ for S10.
 
In the redshift-binned values of Figure \ref{fig:diff3000_differential_final_zbinned} (upper panel), we find that the data has a relatively higher amplitude at lower redshift compared to higher redshift. This brings the data into good agreement with O22 at higher redshift, while maintaining relatively good agreement with Agora and S10 at lower redshift. We also notice strong evolution in O22 between the two sets of redshift-binned values. O22 exhibits significantly higher $\Delta D_{3000}^{yy}$ values at higher redshift compared to lower redshift. By comparing the normalized cluster counts in each mass and redshift bin, we find that this effect is not driven by cosmology. Increases in the individual halo tSZ signal as a function of redshift is not completely unexpected and consistent with self-similar scaling relations found in the literature (e.g., \citealt{Arnaud_2010}). Furthermore, \citet{Shaw_2010} finds that their gas model, which O22 references, is expected to scale close to self-similarly (see Figure 6 in \citealt{Shaw_2010}). The reason Agora and S10 do not scale as strongly with redshift could have something to do with the relatively higher amount of feedback in their gas models \citep{Holder_2001, Battaglia_2012}, but we leave a more in-depth analysis of each gas model for the next sub-section. Ultimately, we note that higher-redshift $\Delta D_{3000}^{yy}$ values from O22 are in better agreement with the data and other simulations.

 In our baseline $\Delta D_{3000}^{yy}$ data values (and in $f_{3000}$), we notice an anomalously high value in the highest-mass bin. However, this discrepancy is not evident in $\Delta D_{3000}^{yy} {\rm \ per \ cluster}$ (discussed in the next subsection). The different simulations may be discrepant with this mean $\Delta D_{3000}^{yy}$ data value for different reasons. In particular, S10 is discrepant due to a lower abundance of highest-mass halos, while Agora is discrepant due to lower individual tSZ cluster signals (as can be seen in the $\Delta D_{3000}^{yy} \ {\rm per \ cluster}$ values of Figure \ref{fig:diff3000_per_cluster}). O22 is discrepant due to both a lower halo abundance and mean individual cluster signal. We could also attribute some of the seemingly high tSZ signal in the data to mass-uncertainties and possible projection effects. Nonetheless, it is important to note that this data point is not very statistically significant and lies within $2\sigma$ of all simulations.

In general, we find the data is broadly consistent with all $\Delta D_{3000}^{yy}$ values from simulations, with a preference for O22 at higher redshift and a preference for Agora and S10 at lower redshift. The main complication is that most of these results are due to a combination of both cosmology and gas model choices. In the next subsection we attempt to disentangle these two components so as to specifically isolate the effects of different gas model prescriptions.

\subsection{$\Delta D_{3000}^{yy} \ {\rm per \ cluster}$}

In Figures \ref{fig:diff3000_per_cluster} and \ref{fig:diff3000_per_cluster_zbinned}, we aim to isolate the effects of gas physics from cosmological and halo-abundance related factors by presenting the difference between masked and unmasked power spectra per cluster. To achieve this, we calculate the ratio of the upper panel values from Figures \ref{fig:diff3000_differential_final} and \ref{fig:diff3000_differential_final_zbinned} to their respective cluster counts in each mass bin, normalized relative to the full sky area. Figure \ref{fig:diff3000_per_cluster} shows results for the full-redshift range, while Figure \ref{fig:diff3000_per_cluster_zbinned} shows the result for our two redshift bins. We note that this metric implicitly assumes that all clusters in a mass bin contribute proportionally the same to the tSZ power spectrum at $\ell = 3000$. 

In the full-redshift limit (Figure \ref{fig:diff3000_per_cluster}), the data shows better agreement with S10 at higher masses, while also demonstrating good agreement with Agora, particularly at lower masses. Overall, the agreement is stronger with Agora, as indicated by a $\chi^{2}$-like value of $6.28$, compared to $10.25$ for S10. O22 exhibits the lowest amplitude and the poorest agreement with the data, with a $\chi^{2}$-like value of $18.81$. 

A striking feature of Figure~\ref{fig:diff3000_per_cluster} is that S10 agrees with the data very well in the highest mass bins but exhibits a shallower slope with mass than the data. The simplest interpretation of this result is that non-thermal pressure support (which S10 lacks) is negligible at the highest mass but important nearer the SPT-SZ completeness limit; however, the literature (e.g., Figure 4 in \citealt{10.1093/mnras/staa1712}) suggests the opposite may be true. An alternative possibility is that non-thermal pressure support is negligible at all masses probed by the data, but that the amount of feedback (e.g., AGN feedback) in S10 is low compared to that in real clusters. AGN feedback is known to have a stronger power-reducing effect on low-mass systems than high-mass systems (e.g., \citealt{Puchwein_2008}; \citealt{2016A&A...592A..46C}), and at our small angular scale of interest (e.g., \citealt{Shaw_2010}). If the amount of feedback were increased in S10, it could be possible to achieve better agreement with the data at low masses while maintaining the current agreement at high masses. However, this could come at the expense of a poorer match to X-ray data.
%The agreement between data and S10 at high mass may be influenced by a number of factors. While a simple interpretation suggests that non-thermal pressure support is less significant at higher masses compared to lower masses, the literature (e.g., Figure 4 in \citealt{10.1093/mnras/staa1712}) suggests the opposite may be true. One alternative possibility is that the amount of feedback (e.g., AGN feedback) in S10 is lower compared to clusters in the data. AGN feedback is known to have a stronger power-reducing effect on low-mass systems than high-mass systems (e.g., \citealt{Puchwein_2008}; \citealt{2016A&A...592A..46C}), and at our small angular scale of interest (e.g., \citealt{Shaw_2010}). If the amount of feedback were increased in S10, it could be possible to achieve better agreement with the data at low masses while maintaining the current agreement at high masses. However, this could come at the expense of a poorer match to X-ray data. Additional investigations and analyses are needed to better understand and reconcile these discrepancies. 

Additionally, we observe that O22 values surpass Agora values at lower masses. This can be attributed to Agora's mass-dependent treatment of the bound gas component in their halo model and their high level of AGN feedback, which has a more significant effect on lower-mass systems as previously discussed. In Section \ref{sec:sims} we describe Agora's calibration approach, including their choice of a gas model with increased AGN feedback and a lower gas-halo mass relation compared to both S10 and O22. We also describe their treatment of the ejected and bound gas fractions in their halo model, which is a distinctive feature of the Agora simulation. The ejected gas component, located outside the halo's virial radius, consists of gas that was initially within the virial radius but has been expelled due to feedback processes, resulting in a lower bound gas fraction. As previously discussed, the fraction of bound gas in Agora increases with mass, such that a higher proportion of relevant tSZ signal comes from high-mass halos. Consequently, we expect low-mass Agora clusters to source a proportionally lower amount of tSZ signal. On the other hand, O22 does not incorporate such a treatment, leading to a less steep decline at lower masses. S10, which excludes non-thermal pressure support and incorporates a lower level of AGN feedback compared to Agora, exhibits higher values overall. This combination of parameter choices helps explain why Agora values mostly lie between S10 and O22 here, yet fall below O22 at lower masses. 

\begin{table}
\centering
\begin{tabular}[t]{ccc}
\toprule 
Product & Best-fit Power-law Index $p$ \\
\hline
SPT-SZ&$3.54 \pm 0.55$\\
S10&$2.50 \pm 0.35$\\
O22&$1.60 \pm 0.12$\\
Agora&$2.61 \pm 0.09$\\
\hline
\end{tabular}
\caption{Best-fit power-law index $p$ (see Equation~\ref{eqn:powerlaw}) and uncertainties for full-redshift $\Delta D_{3000}^{yy} \ {\rm per \ cluster}$ values from Figure \ref{fig:diff3000_per_cluster}. In the case of simulations, only values in the mass range of the data are considered for this calculation.}\label{table:slopes01}
\end{table}

To further examine the mass-dependence, we fit full-redshift data and simulation points to a power-law model:
\begin{equation}
    \Delta D_{3000}^{yy} \ {\rm per \ cluster} = A \ \left ( \frac{M_{500}}{6 \times 10^{14} M_\odot\ h^{-1}} \right )^p
    \label{eqn:powerlaw}
\end{equation}
 and present the best-fit power-law index values $p$ with uncertainties in Table \ref{table:slopes01}. In the case of the simulations, the fit is only performed for values within the same mass range as the data. This ensures a fair comparison between the simulated and observed slopes within the specific mass range of interest. Notably, the data exhibits the steepest index overall, with a value of $3.54 \pm 0.55$. This is followed by Agora, with an index of $2.61 \pm 0.09$, S10, with an index of $2.50 \pm 0.35$, and finally O22, with an index of $1.60 \pm 0.12$. These results highlight the presence of discrepancies between the data and simulations in terms of both the amplitude and mass-dependent scaling.
 
A brief comparison between the best-fit power-law index values obtained for the full-redshift range and those predicted in a self-similar model is worthwhile.  In the self-similar case, the predicted power-law index of the relation between integrated $Y_{500}$ and $M_{500}$ is $5/3$ (e.g., \citealt{Arnaud_2010, 2013}), while the central decrement is predicted to scale linearly with mass ($p=1$, e.g., Equation 7 in \citealt{Holder_2001}). Since we have computed index values based on the tSZ power spectrum, it is necessary to square these self-similar relations such that we instead compare to values of $10/3$  and $2$, respectively. Scaling arguments and simple numerical simulations indicate that, for the typical mass and redshift of clusters considered here, the tSZ power spectrum at $\ell = 3000$ is more sensitive to changes in $Y_{500}$ than the central $y$. In this context, we find that the SPT data exhibits the closest adherence to a self-similar $Y_{500}$ mass-scaling, with a slope value of $3.54 \pm 0.55$. Agora exhibits the next closest adherence to a self-similar mass-scaling, followed by S10, and finally O22. 

In the redshift-binned regime (Figure \ref{fig:diff3000_per_cluster_zbinned}), the data remains in better agreement with S10 and Agora, over O22, particularly at lower redshifts. While higher-redshift data values have larger uncertainties, resulting in a broader range of agreement with all three simulations, the central data values still show closer alignment with S10 at higher masses and Agora at lower masses. Once again, O22 values scale more strongly with redshift, which we would expect from the near self-similar model of \cite{Shaw_2010}. 

Despite O22's relatively closer adherence to a self-similar redshift-scaling, it exhibits the least self-similar mass-scaling among the models. Similarly, we observe a lack of self-similar redshift evolution in Agora and S10, even though these models show closer adherence to a self-similar mass-scaling. Some of these findings could tentatively be attributed to increased feedback and non-gravitational processes, which are predicted to affect redshift-scaling more strongly than mass-scaling \citep{Holder_2001}. Both S10 and Agora incorporate higher levels of feedback in their gas models compared to O22, and Agora includes several other redshift-dependent terms in their gas model (e.g., modifications to halo concentrations) which could collectively contribute to scaling the redshift-evolution of pressure profiles away from self-similarity. Nonetheless, there are clearly other model features or assumptions at play in these results which are not fully understood. Interestingly, the data itself does not exhibit a strong dependence on redshift at $\ell = 3000$. Further analysis and investigation are necessary to fully understand the implications and underlying mechanisms or assumptions driving these discrepancies. 

\begin{figure}
    \centering
    \includegraphics[width=3.3in]{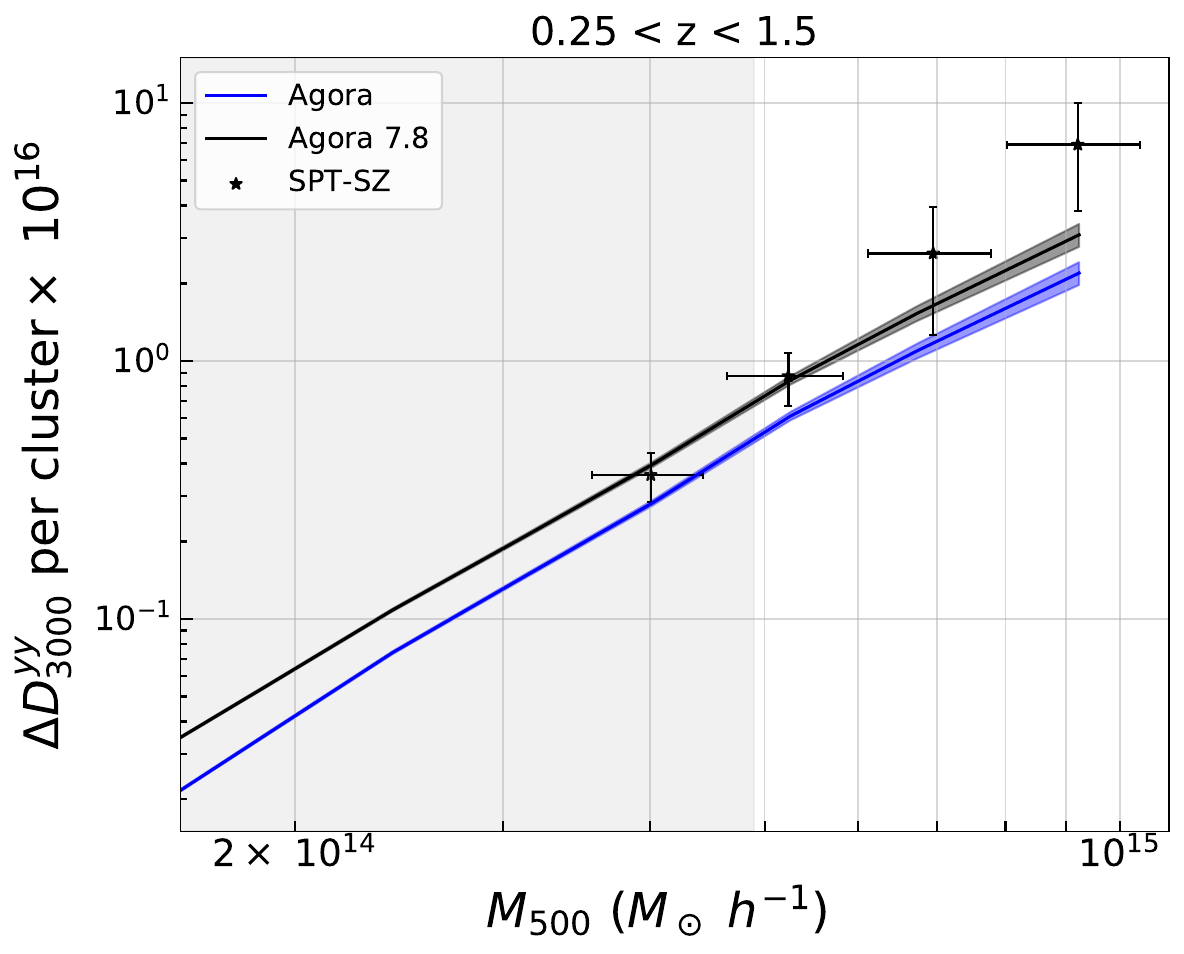}
    \caption{Comparison of $\Delta D_{3000}^{yy} \ {\rm per \ cluster}$ values from an alternative Agora $y$ map, produced using lower levels of AGN feedback, with data and baseline values. Alternative values are labeled as ``Agora $7.8$,'' which refers to the value of the AGN heating temperature parameter in Agora's gas model ($10^{7.8} \, \mathrm{K}$). The baseline Agora $y$ map employs an AGN heating temperature of $10^{8.0} \, \mathrm{K}$ in order to better match overall measurements of the tSZ power spectrum from data \citep{omori2022}. Evidently, we find that reducing the amount of AGN feedback in Agora results in a closer match to the data within the targeted mass range.} 
    \label{fig:d3000_per_cluster_agora78_test}
\end{figure}

The observed differences in amplitude, and in mass and redshift scaling, between simulations and data suggest that the simulated models may not fully capture the underlying physical processes governing the observed data, and some mass- and redshift-dependent adjustments to the gas models would be required to improve the agreement. S10 exhibits discrepancies at lower masses, and potential adjustments to its gas model have already been discussed above. On the other hand, O22 and Agora show larger discrepancies at higher masses, suggesting that these simulations may have overestimated the effects of power-reducing mechanisms in this regime. This seems to be especially true at lower redshift. To address this, reducing the amount of non-thermal pressure support in O22 could be considered, but it would result in a poorer match to the overall amplitude of the tSZ power spectrum from G15. Alternatively, increasing the amount of feedback in O22 could help compensate for the discrepancy and reduce its redshift evolution. In the case of Agora, it is possible that the amount of AGN feedback has been overestimated for the mass range of interest. In Figure \ref{fig:d3000_per_cluster_agora78_test} we compare baseline data and Agora values to those derived from an alternative Agora $y$ map with a lower AGN heating temperature, and consequently less AGN feedback. We find that these alternative values are a better match to the data here, having a $\chi^{2}$-like value of $2.34$, but would yield a worse match to the overall tSZ power spectrum amplitude of G15 and other data (see Figure 15 in \citealt{omori2022}). This suggests overall that feedback processes may not be very significant at high mass while remaining significant at lower masses. Moreover, the data implies a steep mass-dependent scaling that is not reproduced by any of the simulations. 

The $\Delta D_{3000}^{yy}$ metric used here does not capture the tSZ power contributions from low-mass clusters ($M_{500} < 10^{14}\ M_\odot\ h^{-1}$). While S10 agrees well with the data based on this metric, it is expected to be consistently higher in amplitude than the data at lower masses, as indicated by the relatively lower amplitude of the SPT full tSZ power spectrum from G15 (see Figure \ref{fig:all_sim_tsz_ps}). In the next subsection, we incorporate measurements of the full tSZ power spectrum amplitude as an additional probe of low-mass cluster signals. 

\subsection{$f_{3000}$}

Using $f_{3000}$ as a metric for comparison enables us to consider the cumulative impact of various model and data components, including cosmology, gas physics, and the contributions of both high- and low-mass clusters. High-mass clusters are directly examined through our masking routine, while the influence of low-mass clusters is incorporated through the full tSZ power spectrum amplitude, which is primarily sourced by high-redshift low-mass clusters. By considering these factors together, we can gain a comprehensive understanding of the overall agreement between the simulations and the observed tSZ power spectrum.

In the top panel of Figure \ref{fig:f3000_differential_final} we find that the relative position and shape of all points remains largely unchanged compared to their $\Delta D_{3000}^{yy}$ counterparts. However, the mean SPT-SZ data points have a significantly higher relative amplitude in $f_{3000}$. Nevertheless, we find that the data is in good agreement with Agora for most mass bins and overall, yielding a $\chi^{2}$-like value of $3.28$. In comparison, S10 and O22 yield $\chi^{2}$-like values of $6.98$ and $21.52$, respectively. 

In comparing the simulations, Agora consistently exhibits higher amplitudes than the other simulations for mass bins above $M_{500} \sim 3 \times 10^{14}\ M_\odot\ h^{-1}$. This can be attributed to Agora's higher cluster abundance and a relatively lower full tSZ power spectrum amplitude compared to S10. Evidently, there is a higher proportion of tSZ signal at high masses in Agora compared to lowmasses. However, at lower masses, there is a turnover where O22 surpasses both Agora and S10 in amplitude. Despite having a comparable full tSZ power spectrum amplitude to Agora, as shown in Figure \ref{fig:all_sim_tsz_ps}, low-mass clusters in O22 source a proportionally higher amount of tSZ signal. For Agora, this can be partly explained by the mass-dependent treatment of their bound gas fraction and high AGN feedback levels. High-mass clusters in S10 may source a proportionally higher amount of tSZ signal due to the relatively higher amount of feedback, compared to O22.

In the redshift-binned version of $f_{3000}$ (Figure \ref{fig:f3000_differential_final_zbinned}), there is a noticeable discrepancy between the data and simulation values at lower redshift in most mass bins. However, there is better agreement in the higher-redshift bin, with all simulations matching up relatively well to the data. Despite larger discrepancies at lower redshift, Agora still provides the best overall match to the data, likely due to similar reasons as outlined above. 

The discrepancies between the data and simulations in $f_{3000}$, as compared to $\Delta D_{3000}^{yy}$, are due in large part to differences in the full tSZ power spectrum amplitude at $\ell = 3000$. This indicates that either the simulations overpredict the tSZ power coming from low-mass clusters, or there is something (e.g., the CIB) contaminating the tSZ signal from low-mass clusters in the data. In Section \ref{subsec: CIB_test} of this work, we find that the CIB does not significantly contaminate high-mass clusters in our baseline data products. However, it is possible that CIB contamination could be more prominent in low-mass clusters, which are more abundant at higher redshift due to hierarchical structure formation. Although G15 attempts to account for CIB contamination by incorporating a simple non-physical model of tSZ-CIB correlation into their fitting procedure (see Equation 14 in G15), this model may not adequately describe the CIB at low masses and in the $\ell$ range of interest. Overall, the differences in gas model choices, binned cluster counts, and full tSZ power spectrum amplitude all play a role in the observed discrepancies in $f_{3000}$. However, it is the variations in the full tSZ power spectrum values that are particularly significant in highlighting larger discrepancies in the low-mass regime.

\begin{table*}
\centering
\begin{tabular}{l|l|l} \hline
\multicolumn{1}{l|}{\textbf{Simulation}}&\multicolumn{1}{l|}{\textbf{Agreement with Data}}&\multicolumn{1}{l}{\textbf{Potential Interpretations}}\\\hline
\textbf{S10}&\begin{minipage}{5cm} ~\\ $\Delta D_{3000}^{yy} \ {\rm per \ cluster}$ at higher \\ masses. $\Delta D_{3000}^{yy}$ at lower \\ redshift. \\ \end{minipage} & \begin{minipage}{5cm} ~\\ Simulated clusters at higher mass and lower redshift \\ are a closer match to the data.\\ More non-thermal pressure \\ support and/or feedback is needed.
\end{minipage} \\\hline
\textbf{O22}&\begin{minipage}{5cm} ~\\ $\Delta D_{3000}^{yy}$ at higher redshift, $f_{3000}$ at higher redshift. \\ \end{minipage} & \begin{minipage}{5cm} ~\\Simulated clusters at higher \\ redshift are a better match to data. \\ Amount of non-thermal pressure \\and stellar-mass-fraction is \\ potentially too high. \\
\end{minipage} \\\hline
\textbf{Agora}&\begin{minipage}{5cm} ~\\ $\Delta D_{3000}^{yy}$ at lower \\ redshift, $\Delta D_{3000}^{yy} \ {\rm per \ cluster}$ \\ at lower masses, \\ full-redshift $f_{3000}$. \\ \end{minipage} & \begin{minipage}{5cm} ~\\ Simulated clusters at lower mass \\ and redshift are a closer match \\ to data. AGN feedback is too \\ strong at higher masses. \\
\end{minipage} \\\hline
\end{tabular}
\caption{Qualitative summary of agreement between simulations and data in this work along with potential interpretations. More details available in the text, along with a brief discussion of possible contaminants in the data.} \label{agreements_between_sims_&_data}
\end{table*}

\section{Robustness Tests}

\label{sec:robustness_tests}
In this section we outline several tests for evaluating the robustness of our baseline results.

\subsection{Cross-check with Literature}
\label{G15_comparison}
In this subsection we compare the values we estimate for $\Delta D_{3000}^{yy}$ in SPT-SZ data to the results of a similar procedure performed in G15.

In addition to their baseline tSZ power spectrum amplitude, G15 reports a value for the tSZ power spectrum with clusters masked. For their version of this calculation, G15 masks all clusters in the 2500 deg$^2$ SPT-SZ survey that are included in the \citet{Bleem_2015} catalog with a statistical significance of $\xi > 5$ (where $\xi$ is the raw cluster signal-to-noise). In their treatment, clusters are masked using apodized masks with $R_{\rm tophat} = 5'$ and a Gaussian taper of ${\rm FWHM} = 5'$ for each. This is notably different from our baseline masking routine. G15 obtains a value for the difference between the masked and unmasked power spectrum, of $\Delta D_{3000}^{yy} = 2.54 \times 10^{-13}$ in units of Compton-$y$. It is difficult to approximate the uncertainty on this measurement due to the correlation between the uncertainties on the masked and unmasked $D_{3000}^{yy}$ measurements from G15. However, we note that it is likely comparable to the uncertainty on our own measurement (reported below). 

For comparison, we reproduce this measurement using our baseline masking routine. That is, we mask clusters with $\xi > 5$ in \citet[][which is an updated catalog of \citealt{Bleem_2015}, but with the same $\xi$ values]{Bocquet_2019} using non-apodized masks, with $\theta_{500}$-dependent radii. We obtain a value of $\Delta D_{3000}^{yy} = (2.75 \pm 0.26) \times 10^{-13}$, which is in good agreement with G15, especially considering the differences in masking routines. 

\subsection{Different Masking Routine Tests}
In this subsection we analyze the effects of altering our masking routine on $\Delta D_{3000}^{yy}$ data values. Specifically, we test the effects of changing the radius prescription of cluster masks and of apodizing masks with Gaussian tapers.

\begin{figure}
    \centering
    \includegraphics[width=3.3in]{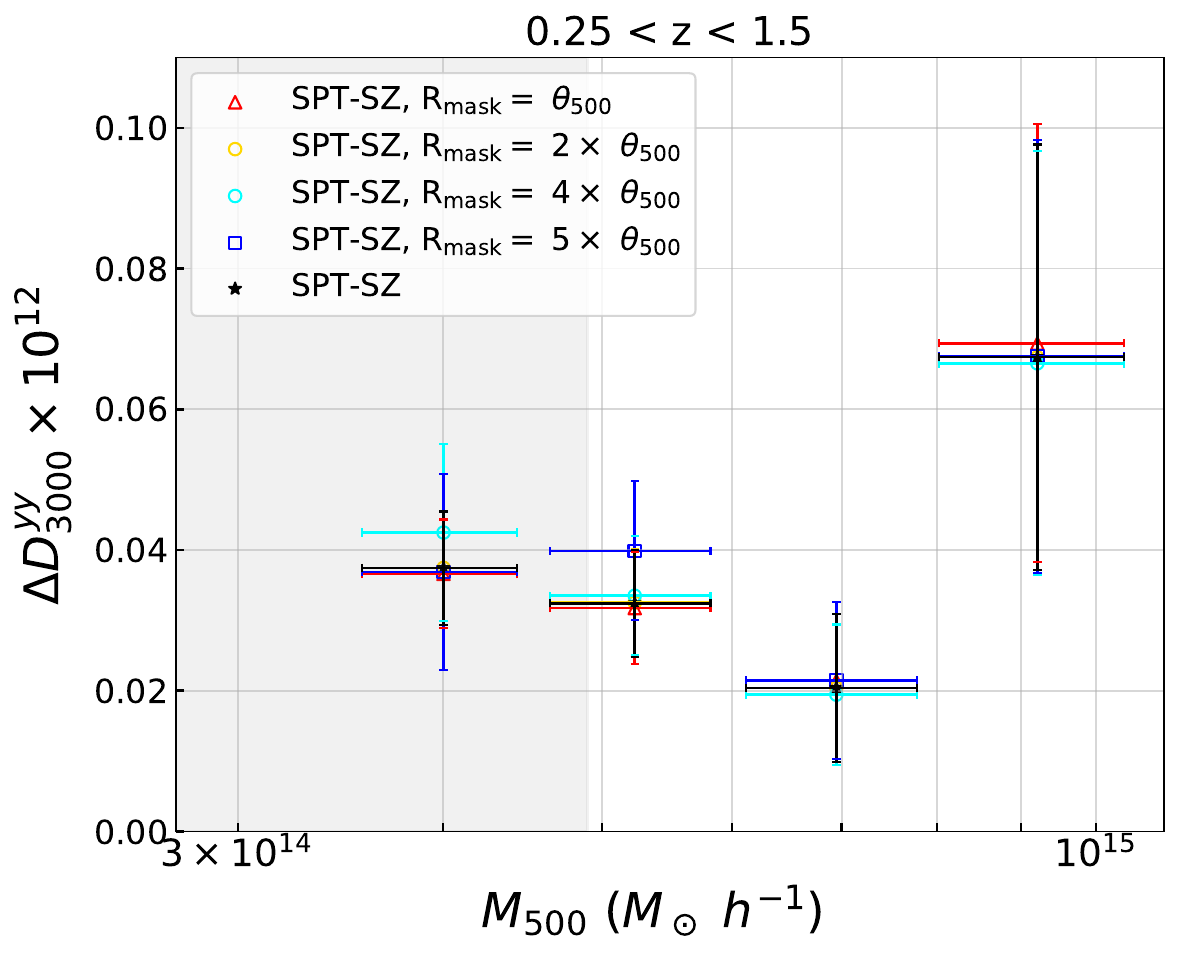}
    \caption{Four sets of $\Delta D_{3000}^{yy}$ data points derived using different mask radii than employed in our baseline $\Delta D_{3000}^{yy}$ (also shown). The legend specifies the different mask radii in units of $\theta_{500}$. For these tests, and unlike in our baseline routine, we do not apply a mask radius floor for clusters with $\theta_{500} < 2'$. Our results appear to be robust across different mask radii choices, as they all yield statistically consistent outcomes.} 
    \label{fig:Diff_size_masking_radii}
\end{figure}

In Figure \ref{fig:Diff_size_masking_radii} we show four sets of $\Delta D_{3000}^{yy}$ data points with different $\theta_{500}$-dependent mask radius functions (as specified in the legend), alongside baseline data points. We notice that these all yield statistically consistent results. This indicates that our baseline  data values are not sensitively dependent on our choice of cluster mask radius.

\begin{figure}
    \centering
    \includegraphics[width=3.3in]{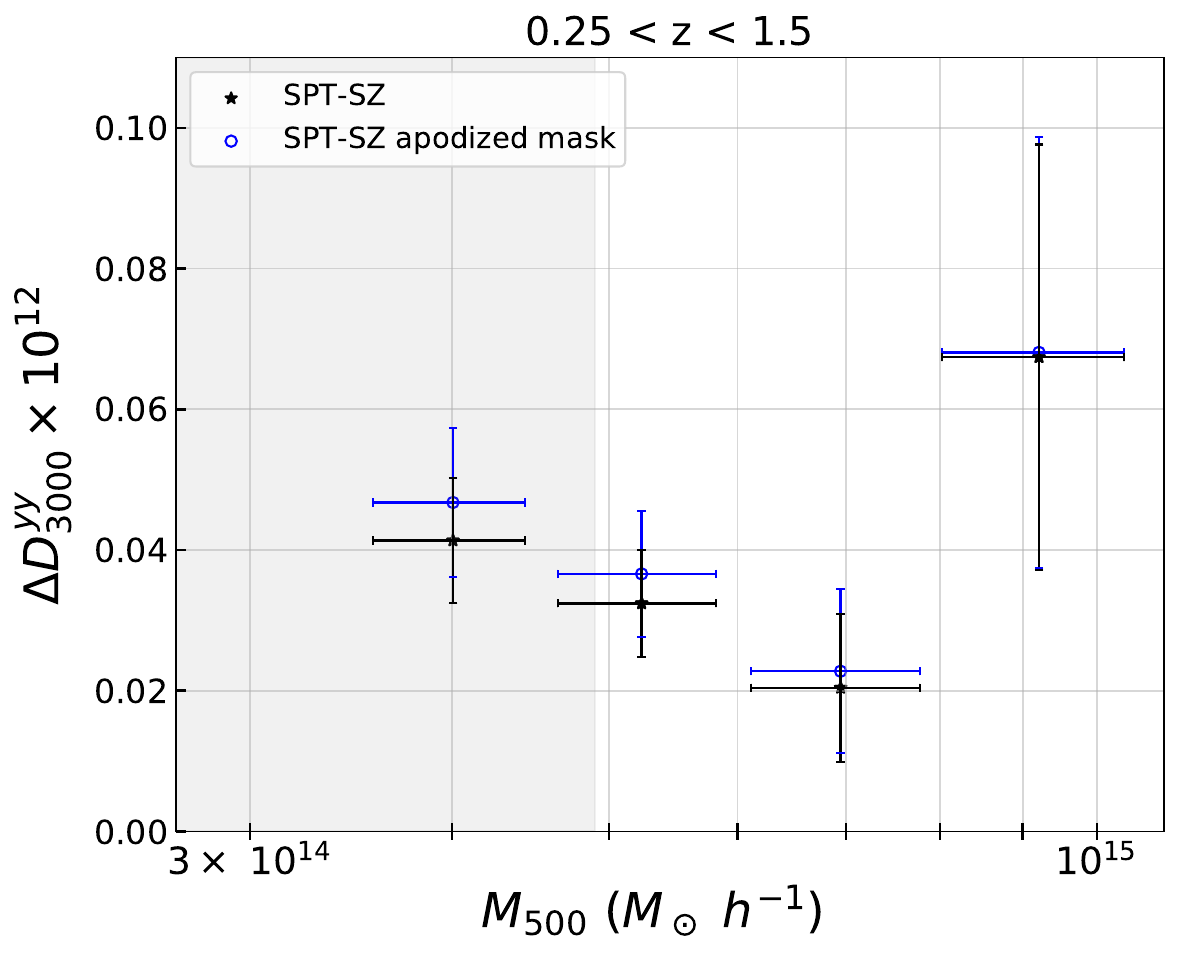}
    \caption{$\Delta D_{3000}^{yy}$ values obtained using Gaussian apodized masks (blue circle markers), alongside baseline values (black star markers). For the apodized masks we apply our baseline masking routine with the addition of a Gaussian taper, defined by ${\rm FWHM} = 5'$, to all masks. This is mostly a test for any significant mode-coupling effects produced by our tophat masking routine. Our results show no indication of such significant mode-coupling effects.} 
    \label{fig:gauss_apod_plots}
\end{figure}

In Figure \ref{fig:gauss_apod_plots} we present $\Delta D_{3000}^{yy}$ data points created using apodized masks, alongside baseline data points. For the apodized masks we apply our baseline masking routine with the addition of a Gaussian taper, defined by ${\rm FWHM} = 5'$, to all masks. This is mostly a test for any significant mode-mixing effects produced by our tophat masking routine. We generally find that the mean data points associated with apodized masks are higher in $\Delta D_{3000}^{yy}$ amplitude than our baseline points. This might indicate that more residual tSZ signal from filamentary structure, or nearby halos, is being masked and is not necessarily due to mode-mixing effects. Nevertheless, we find that the difference between both sets of data points is not very significant (< 1$\sigma$) and adopting the test values would not change our overall discussion and conclusion.

\subsection{Comparison to ACT Compton-$y$ Map}

\begin{figure}
    \centering
    \includegraphics[width=3.3in]{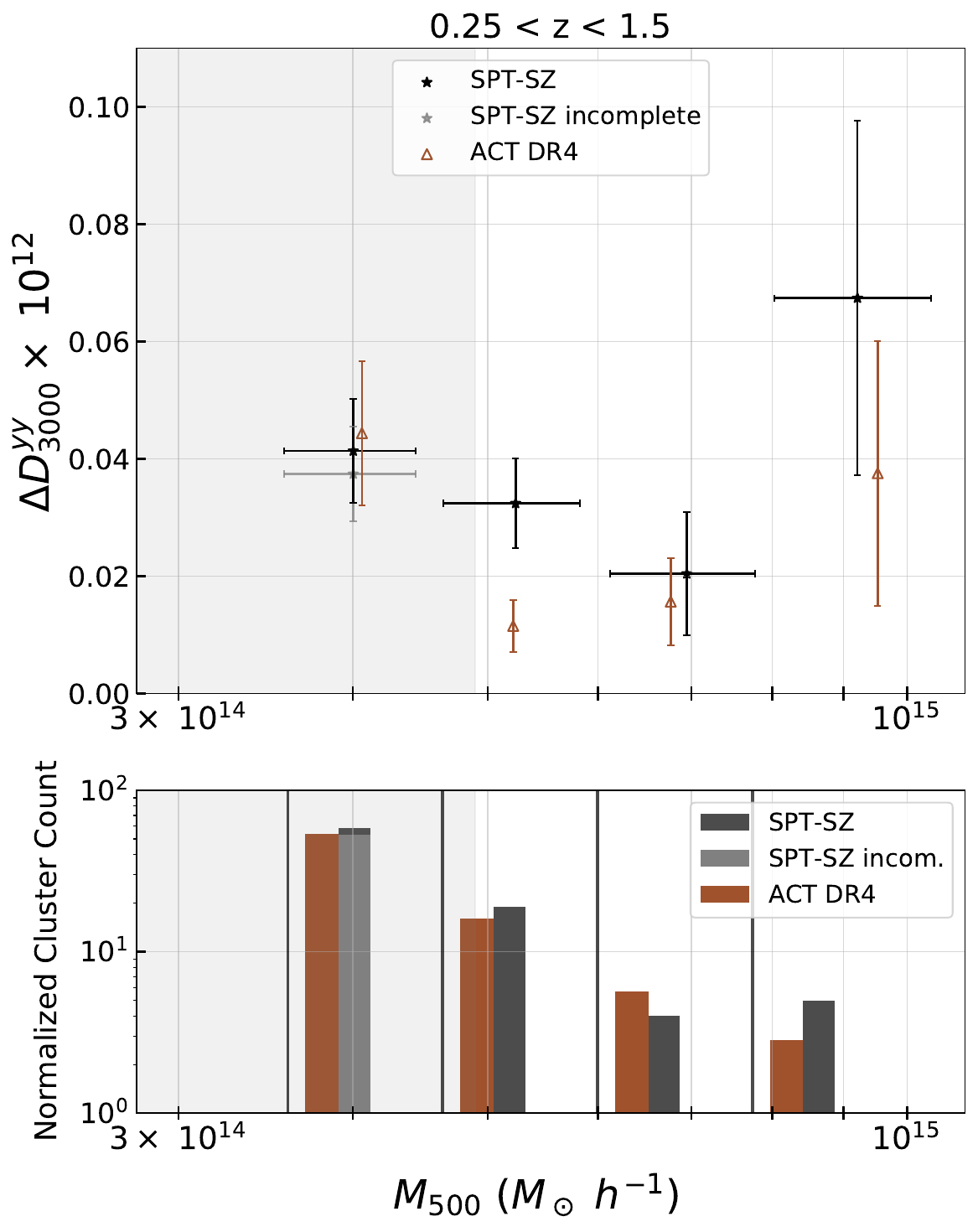}
    \caption{\textit{Top panel}: $\Delta D_{3000}^{yy}$ values derived from an ACT DR4 Compton-$y$ map alongside baseline SPT-SZ data values. There is generally decent agreement between ACT and SPT data points here. \textit{Bottom panel}: A histogram of cluster counts in the ACT and SPT-SZ catalogs, normalized relative to the SPT-SZ sky area.} 
    \label{fig:ACT_comparison}
\end{figure}

\begin{figure}
    \centering
    \includegraphics[width=3.3in]{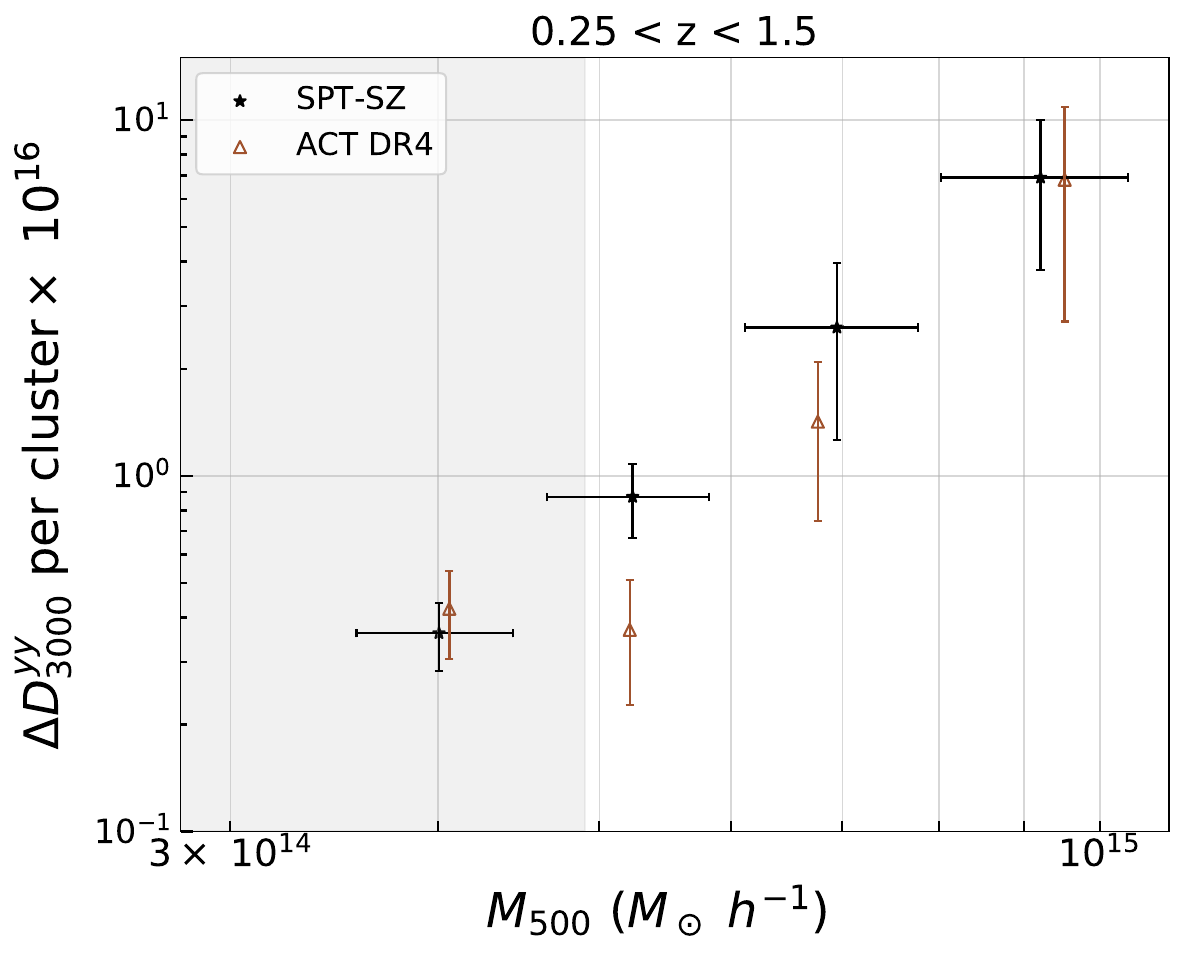}
    \caption{$\Delta D_{3000}^{yy} \ {\rm per \ cluster}$ values derived from an ACT DR4 Compton-$y$ map alongside baseline SPT-SZ data values. We find good agreement overall between ACT and SPT data points here as well.} 
    \label{fig:ACT_comparison_per_cluster}
\end{figure}

 As a cross-check, we wish to compare baseline SPT-SZ data values to those derived from a different data product. In the top panel of Figure \ref{fig:ACT_comparison} we compare SPT-SZ $\Delta D_{3000}^{yy}$ measurements to those derived from the Compton-$y$ map described in \citet{2020PhRvD.102b3534M}, which was produced from the combination of data from the ACTPol receiver on the Atacama Cosmology Telescope (ACT, \citealt{2016ApJS..227...21T}) and the \textit{Planck} satellite. The bottom panel includes a normalized histogram of cluster counts in each bin. In Figure \ref{fig:ACT_comparison_per_cluster} we compare $\Delta D_{3000}^{yy} \ {\rm per \ cluster}$ values for both data products. In order to measure these metrics for ACT clusters, we use the cluster catalog described in \citet{2021ApJS..253....3H} and apply our baseline masking and $\Delta D_{3000}^{yy}$-error-approximation routine. We note that the mass-significance scaling relation of ACT is different than that of SPT-SZ, and so the cluster masses in each catalog are not directly comparable to each other. We attempt to compensate for this by scaling all ACT halo masses by a value of $1.027$, the estimation of which is described in Section 5.1 of \citet{2021ApJS..253....3H}. Additionally, we have omitted error approximations on ACT mass values for simplicity.
 
In Figure \ref{fig:ACT_comparison}, we generally notice that most ACT $\Delta D_{3000}^{yy}$ data values are in agreement with our baseline SPT-SZ results. There is similarly high agreement in $\Delta D_{3000}^{yy} \ {\rm per \ cluster}$, as seen in Figure \ref{fig:ACT_comparison_per_cluster}. 

\subsection{Effects of CIB Contamination}
\label{subsec: CIB_test}

\begin{figure}
    \centering
    \includegraphics[width=3.3in]{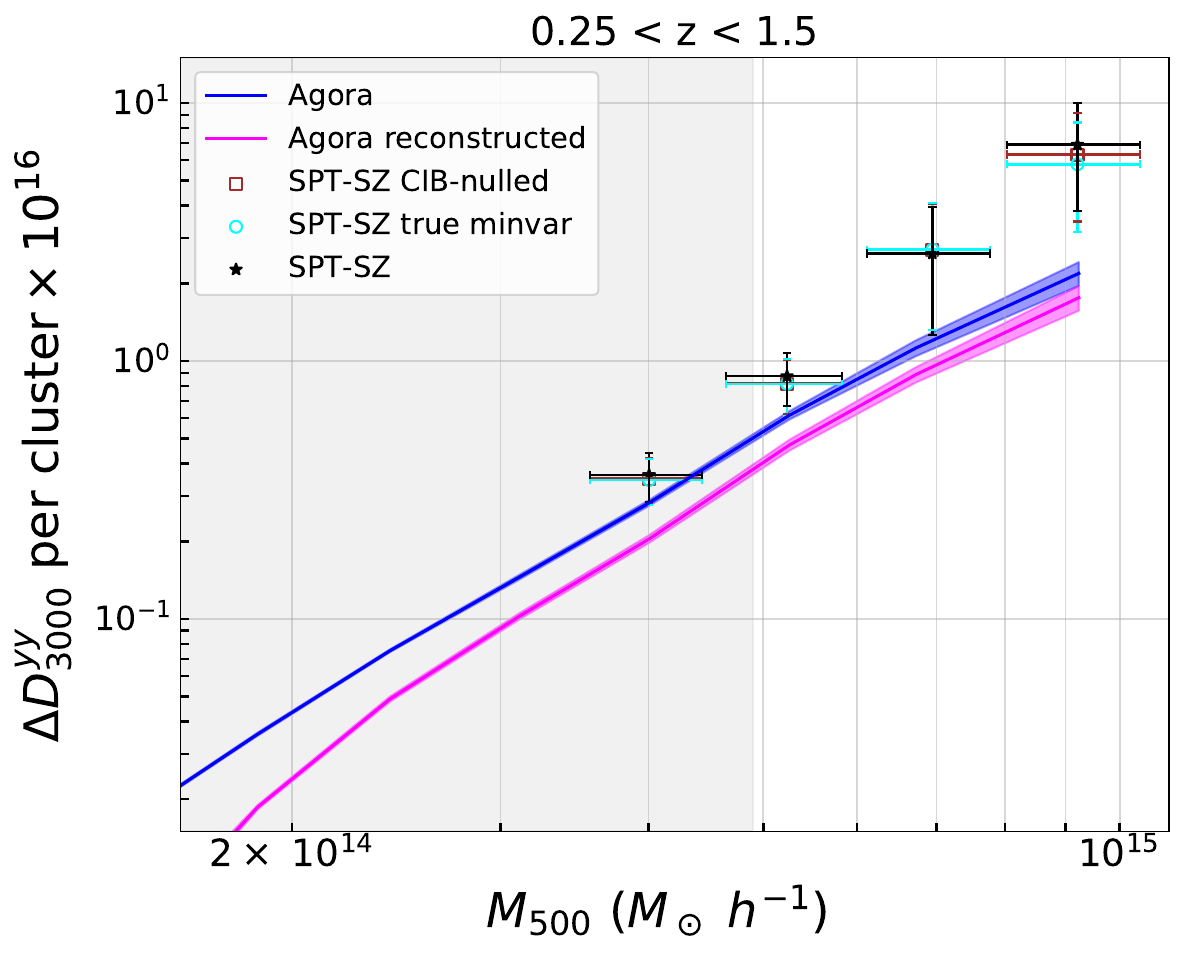}
    \caption{$\Delta D_{3000}^{yy} \ {\rm per \ cluster}$ values derived from different data and simulation products in order to test for the significance of CIB-contamination on our baseline values (also shown). Brown square markers, labeled ``SPT-SZ CIB-nulled,'' are obtained using an SPT-SZ $y$ map with a single CIB component explicitly nulled. Cyan circle markers, labelled ``SPT-SZ true minvar,'' are derived using an SPT-SZ $y$ map which does not include any mitigation for CIB contamination. Baseline $\Delta D_{3000}^{yy} \ {\rm per \ cluster}$ values for Agora are also shown, and compared with values from a reconstructed Agora $y$ map which includes contamination from a model of the CIB. This reconstructed map is created using the same $y$ map frequency weights used to create the $y$ map we use for our baseline results. Notably, there is no indication of significant CIB contamination in our targeted mass and redshift range.} 
    \label{fig:cib_test}
\end{figure}

\begin{figure*}
    \centering
    \includegraphics[width=7.1in]{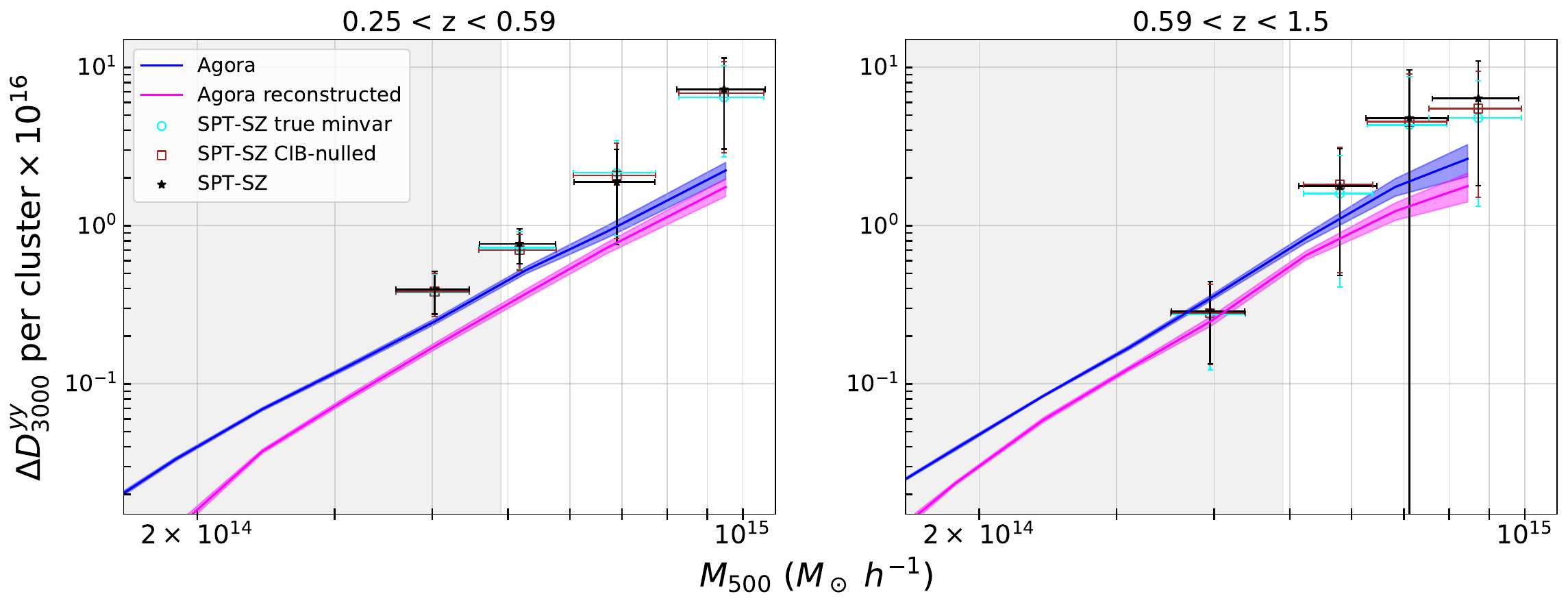}
    \caption{ $\Delta D_{3000}^{yy} \ {\rm per \ cluster}$ values similar to Figure \ref{fig:cib_test} but for our two redshift bins. There is no indication of significant CIB contamination in our targeted mass and redshift range.} 
    \label{fig:cib_test_zbinned}
\end{figure*}

In our high-$\ell$ region of interest, the CIB is expected to be the primary source of microwave contamination in our SPT-SZ Compton-$y$ map, especially after masking out radio point sources (see Figure 2 in \citealt{2021ApJ...908..199R}). We run three separate tests probing the effects of CIB contamination on our baseline $\Delta D_{3000}^{yy} \ {\rm per \ cluster}$ values and show the results in Figures \ref{fig:cib_test} (full-redshift) and \ref{fig:cib_test_zbinned} ($z$-binned). We choose $\Delta D_{3000}^{yy} \ {\rm per \ cluster}$ as a metric for comparison for this test so as to factor out the effects of different cluster counts between data and simulations.

The first two tests involve running two alternate SPT-SZ Compton-$y$ maps, with different CIB treatments, through our analysis pipeline. The data values labeled ``SPT-SZ CIB-nulled'' in the legend of Figures \ref{fig:cib_test} and \ref{fig:cib_test_zbinned} are obtained using an SPT $y$ map created using a linear combination algorithm that nulls a particular CIB spectral energy distribution. Recall that our baseline results (labeled ``SPT-SZ'') use a $y$ map that has had the weights slightly altered to reduce (though not explicitly null) the CIB, assuming a particular two-component model. Finally, the data points labelled as ``SPT-SZ true minvar,'' were derived using an SPT-SZ $y$ map which does not include any mitigation of CIB contamination beyond simply minimizing the total noise-plus-foreground variance. We note that the alternate $\Delta D_{3000}^{yy} \ {\rm per \ cluster}$ values are nearly identical to our baseline results. This indicates that our data is not significantly contaminated by the CIB in our high-mass range of interest.

As an additional test, we compute $\Delta D_{3000}^{yy} \ {\rm per \ cluster}$ values for a reconstructed Agora $y$ map and also present these results in Figures \ref{fig:cib_test} and \ref{fig:cib_test_zbinned}. This particular product is created by %incorporating a model of the CIB, as well as the CMB, and then 
taking the per-frequency maps from Agora (which contain all sky components, including tSZ, primary CMB, CIB, and radio sources) and
reconstructing the $y$ map using the same frequency weights used to create our baseline SPT-SZ $y$ map (for more details see \citealt{Bleem_2022} and \citealt{omori2022}).
Based on these alternate values, it seems that the main effect of including CIB contamination in the Agora $y$ map is to cause an overall decrease in $\Delta D_{3000}^{yy} \ {\rm per \ cluster}$, as compared to baseline Agora values. Since our tests on the data do not suggest the presence of significant levels of CIB contamination, we believe the Agora reconstructed $y$ map likely overapproximates the effects of CIB contamination at the masses tested here.

\section{Conclusion}
\label{sec:conclusion}

Measurements of the tSZ effect have the potential to constrain cosmological parameters to exquisite precision. However, such constraints are difficult to achieve in practice due to modelling limitations. In this work we study features of the tSZ power spectrum as a function of halo mass and redshift, utilizing semi-analytical $N$-body simulations and SPT-SZ data, with the intention of analyzing and constraining different models of the ICM.

In general, we find notable discrepancies between the data and all ICM gas models considered, suggesting the need for additional mass-dependent adjustments to each model, with caveats and limitations outlined in Section \ref{sec:summary_and_discussion}. Notably, the data predicts a steep mass-scaling which all of the simulations fail to capture. Nonetheless, the gas models of S10 and Agora provide the best fit to data in $\Delta D_{3000}^{yy} \ {\rm per \ cluster}$, with S10 providing a better fit at the highest halo masses and Agora a better fit at lower masses. O22 is very broadly consistent with the data in full-redshift $\Delta D_{3000}^{yy}$ and in all metrics at higher redshift, but generally in worst agreement overall. 

The agreement between the data and S10, as well as the agreement between data and an alternative low-AGN-feedback Agora $y$ map (Figure \ref{fig:d3000_per_cluster_agora78_test}), indicates that the role of feedback is less significant at high mass (masses tested here). However, at low masses ($M_{500} < 10^{14}\ M_\odot\ h^{-1}$), AGN feedback must play a significant role to achieve better agreement with data measurements of the full tSZ power spectrum. It is also possible that the implied discrepancy between the data and all simulations at low mass is due to contamination by sources such as dusty galaxy emission in the tSZ power spectrum measured in the data. These sources could correspond to clusters with masses which are below the SPT-SZ cluster selection threshold but which source a large fraction of the tSZ power at $\ell=3000$. If we assume the measured power is uncontaminated, the combination of Figure \ref{fig:all_sim_tsz_ps} and full-redshift $f_{3000}$ suggest that the Agora baseline model simulates low mass cluster signals in better agreement with the data than S10 or O22. Despite the aforementioned discrepancies, Agora provides the best overall fit to the data as it remains statistically consistent within $1.5 \sigma$ and yields the lowest $\chi^{2}$-like values across all metrics considered. 

Furthermore, the data does not suggest the need for any significant evolution in individual tSZ signals (at $\ell = 3000$) between the two redshift bins used here. This is seemingly a departure from the redshift dependence in the simplest self-similar scenario. Although there are notable differences between $z$-binned data values in $f_{3000}$ and $\Delta D_{3000}^{yy}$, this is mainly due to differences in relative cluster counts between our two redshift bins. The agreement between $\Delta D_{3000}^{yy} \ {\rm per \ cluster}$ values in the two $z$ bins indicates that the data gas pressure signal does not evolve strongly with redshift. It is worth noting that among the simulations considered, only O22 includes a strong redshift evolution and generally models higher redshift clusters in better agreement with the data compared to lower-redshift clusters. 
   
While the analysis presented here provides some valuable insights, it only begins to scratch the surface of potential comparisons between data and simulations. A more complete cluster catalog and statistically significant set of data points would be helpful in making more robust statements and comparisons. This should soon be possible with data from SPT-3G \citep{Sobrin_2022}, Simons Observatory \citep{2019JCAP...02..056A}, and CMB-S4 \citep{abazajian2019cmbs4}. Furthermore, there exist a plethora of other features of the tSZ effect and its power spectrum that could be used as a basis of comparison between models, such as the shape or $\ell$-dependence as a function of mass and redshift, to name two. Despite limitations, we hope this paper will inspire further analysis and development of semi-analytical $N$-body simulations, which undoubtedly are invaluable tools in our understanding of cosmology and cluster physics.

\section{Acknowledgements}
The authors thank Dhayaa Anbajagane for helpful comments and suggestions.

The South Pole Telescope program is supported by the National Science Foundation through the awards OPP-1852617 and OPP-2147371. Partial support is also provided by the Kavli Institute of Cosmological Physics at the University of Chicago.
Argonne National Laboratory's work was supported by the U.S. Department of Energy, Office of High Energy Physics,
under Contract No. DE-AC02-06CH11357. 

The CosmoSim database used in this paper is a service by the Leibniz-Institute for Astrophysics Potsdam (AIP). The MultiDark database was developed in cooperation with the Spanish MultiDark Consolider Project CSD2009-00064.

The authors gratefully acknowledge the Gauss Centre for Supercomputing e.V. (www.gauss-centre.eu) and the Partnership for Advanced Supercomputing in Europe (PRACE, www.prace-ri.eu) for funding the MultiDark simulation project by providing computing time on the GCS Supercomputer SuperMUC at Leibniz Supercomputing Centre (LRZ, www.lrz.de). The Bolshoi simulations have been performed within the Bolshoi project of the University of California High-Performance AstroComputing Center (UC-HiPACC) and were run at the NASA Ames Research Center.

%\vspace{5mm}
%\nocite{apsrev41Control}
\bibliographystyle{yahapj}
\bibliography{manuscript}
\end{document}